\documentclass[
 prxquantum,twocolumn,
 amsmath,amssymb,
 aps,nofootinbib,nolongbibliography,
]{revtex4-2}

\usepackage{amsthm}
\usepackage{graphicx}
\usepackage{color}
\usepackage{dcolumn}
\usepackage{bm}
\usepackage{mathtools}      
\usepackage{mathrsfs}       
\usepackage{braket}
\usepackage{float}
\usepackage[table]{xcolor}

\newcommand{\bx}{\ensuremath{\boldsymbol{x}}}
\newcommand{\by}{\ensuremath{\boldsymbol{y}}}

\newcommand{\br}{\ensuremath{\boldsymbol{r}}}

\newcommand{\bR}{\ensuremath{\boldsymbol{R}}}

\newcommand{\demb}{\ensuremath{d}_\textup{emb}}

\DeclareMathAlphabet{\mymathbb}{U}{BOONDOX-ds}{m}{n}
\usepackage[utf8]{inputenc} 

\usepackage{hyperref}
\hypersetup{
    colorlinks = true,
    linkcolor = {blue},
    citecolor = {blue},
    urlcolor = {blue},
}

\usepackage{titlesec}
\titleformat*{\section}{\raggedright\large\bfseries\sffamily}
\titleformat*{\subsection}{\raggedright\bfseries\sffamily}

\begin{document}
\title{Molecular Quantum Transformer}

\author{Yuichi Kamata}
\email{kamata.yuichi@fujitsu.com}
\affiliation{Quantum Laboratory, Fujitsu Research, Fujitsu Limited, Kawasaki, Kanagawa 211-8588, Japan}

\author{Quoc Hoan Tran}
\email{tran.quochoan@fujitsu.com}
\thanks{Corresponding author}
\affiliation{Quantum Laboratory, Fujitsu Research, Fujitsu Limited, Kawasaki, Kanagawa 211-8588, Japan}

\author{Yasuhiro Endo}
\affiliation{Quantum Laboratory, Fujitsu Research, Fujitsu Limited, Kawasaki, Kanagawa 211-8588, Japan}

\author{Hirotaka Oshima}
\affiliation{Quantum Laboratory, Fujitsu Research, Fujitsu Limited, Kawasaki, Kanagawa 211-8588, Japan}


\begin{abstract}
The Transformer model, renowned for its powerful attention mechanism, has achieved state-of-the-art performance in various artificial intelligence tasks but faces challenges such as high computational cost and memory usage. Researchers are exploring quantum computing to enhance the Transformer's design, though it still shows limited success with classical data. With a growing focus on leveraging quantum machine learning for quantum data, particularly in quantum chemistry, we propose the Molecular Quantum Transformer (MQT) for modeling interactions in molecular quantum systems. By utilizing quantum circuits to implement the attention mechanism on the molecular configurations, MQT can efficiently calculate ground-state energies for all configurations. Numerical demonstrations show that in calculating ground-state energies for $\textup{H}_{2}$, LiH, $\textup{BeH}_{2}$, and $\textup{H}_{4}$, MQT outperforms the classical Transformer, highlighting the promise of quantum effects in Transformer structures. Furthermore, its pretraining capability on diverse molecular data facilitates the efficient learning of new molecules, extending its applicability to complex molecular systems with minimal additional effort. Our method offers an alternative to existing quantum algorithms for estimating ground-state energies, opening new avenues in quantum chemistry and materials science.

\end{abstract}
\maketitle


\begin{figure*}
	\includegraphics[width=18cm]{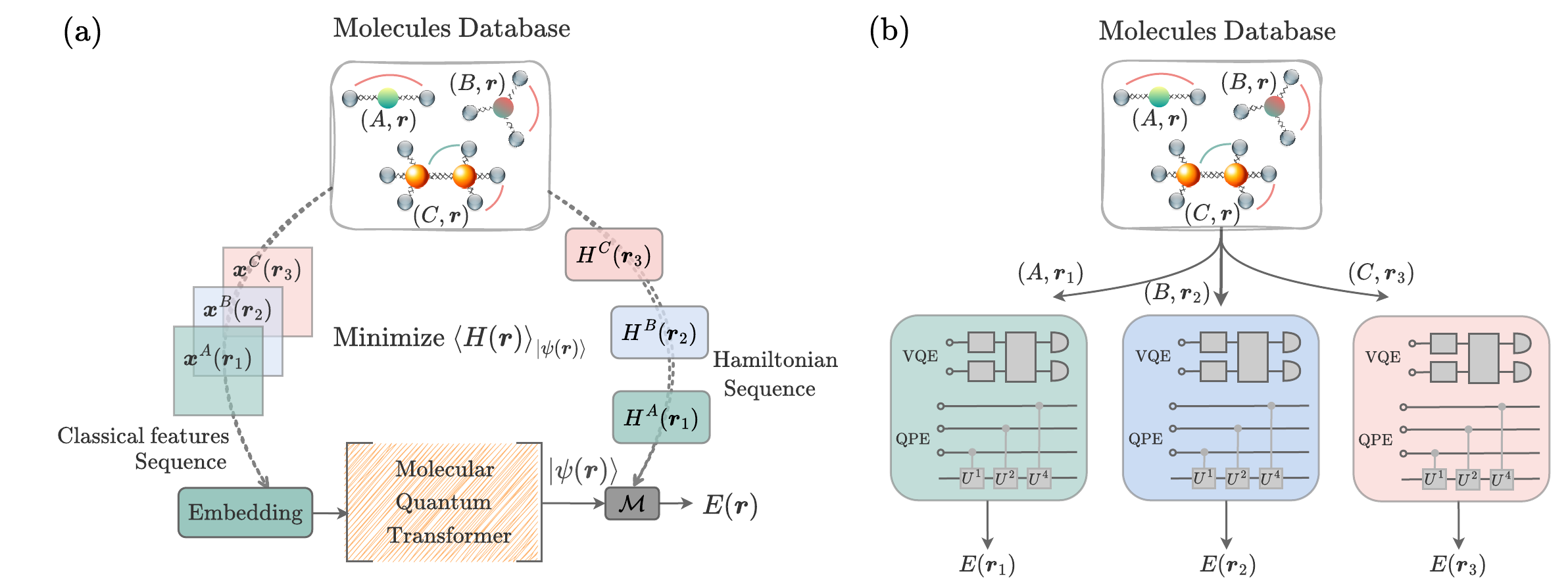}
	\protect{\caption{Overview of the Molecular Quantum Transformer (MQT) model for the ground-state energy calculation across various molecules and their configurations, and the comparison with traditional methods. (a) For each molecule A, B, C,\ldots and its associated configuration $\br_1, \br_2, \br_3, \ldots$, the MQT receives a corresponding classical features sequence $\bx^A(\br_1), \bx^B(\br_2), \bx^C(\br_3), \ldots$ through an embedding process. Leveraging a quantum attention mechanism, the MQT represents the complex interactions and correlations within the molecular system. The output of the MQT is a sequence of quantum states $\ket{\psi^A(\br_1)}, \ket{\psi^B(\br_2)}, \ket{\psi^C(\br_3)}, \ldots$, which reflects these correlations in a variational representation of the estimated ground states for $(A, \br_1), (B, \br_2), (C, \br_3), \ldots$, respectively. The corresponding Hamiltonian $H^A(\br_1), H^B(\br_2), H^C(\br_3), \ldots$, derived from quantum mechanics are transformed into measurable operators to be measured on $\ket{\psi^A(\br_1)}, \ket{\psi^B(\br_2)}, \ket{\psi^C(\br_3)}, \ldots$. During training, the optimization process adjusts the variational parameters in both the MQT and the embedding process to minimize the expectation value $\braket{H(\br)}_{\ket{\psi{(\br)}}}$  across various molecules and a range of $\br$ values. In the evaluation phase, given a molecule, the MQT can provide an estimator of ground-state energy $E(\br)$ for any configuration $\br$. (b) In contrast,  traditional methods such as VQE or QPE require an independent and computationally expensive solver for each molecule and configuration $\br$.\label{Figure:overview}}}
\end{figure*}

The Transformer model~\cite{attention:2017:NIPS} has been recognized as a remarkable advancement in artificial intelligence. Its key power lies in its ``attention mechanism", which discerns the relative importance of different parts of its input and the connection strengths between them. This mechanism has been successfully applied to both natural language processing and visual object recognition tasks, delivering state-of-the-art performance across a variety of datasets. Despite these successes, the current implementation of the Transformer faces several challenges, including high computational costs, substantial memory requirements, the necessity for large datasets, and a vast number of training parameters. These limitations have prompted researchers to explore improved Transformer designs.

The marriage of quantum computing and machine learning has given rise to the field of quantum machine learning (QML)~\cite{dunjko:2016:prl:qmlenhanced,biamonte:2017:QML,dunjko:2020:nonreview,schuld:2021:qmlbook,halvlicek:2019:supervised,schuld:2019:feature}, which aims to leverage quantum computers to tackle problems that are infeasible for classical computers. In this context, efforts have been made to develop quantum versions of the Transformer model.
A significant advancement in quantum neural networks (QNNs) involves integrating the self-attention mechanism by encoding query and key vectors as quantum states using parametric quantum circuits (PQCs). This adaptation enables quantum analogs of the classical self-attention framework, though various approaches differ in their implementation. One straightforward extension replaces the classical inner-product self-attention with the overlap of quantum states~\cite{disipio:2021:QNLP}, but this method struggles to scale effectively for capturing correlations in large datasets due to its computational complexity. To address this limitation, an alternative employs classical Gaussian projections of query and key quantum states, enhancing scalability while retaining essential features~\cite{li:2024:quasn}. Other quantum self-attention variants include quantum vision Transformers~\cite{xue:2024:quantumvision} as an end-to-end approach that leverages analog encoding with quantum random access memory (qRAM)~\cite{lloyd:2008:QRAM}, and hybrid classical-quantum methods~\cite{cherrat:2024:quantumvision,smaldone:2025:hybridtrans} to reduce the time complexity of computing query-key dot products. Additionally, some proposals diverge from inner-product-based attention entirely, opting instead to mix tokens directly in Hilbert space to model correlations without explicitly calculating query-key dot products~\cite{zheng:2023:QSAN,evans:2025:qat}. Despite these innovative adaptations, these quantum self-attention implementations have demonstrated limited performance on tasks involving classical data, such as text and image classification, highlighting challenges in translating the advantages of quantum Transformer to practical applications.

In recent years, there has been a growing recognition within the research community that classical data, such as text and images, do not inherently require quantum effects for processing. Consequently, there is a shift towards employing QML methods that exploit quantum effects on data originating from quantum systems~\cite{editorial:2023:qmladv:nature}. One promising yet underexplored area is the application of QML methods in quantum chemistry, where QML holds the potential to determine molecular and material properties more efficiently than traditional quantum computational chemistry algorithms.

A central focus in quantum computational chemistry is the electronic structure problem, which involves calculating the electronic Hamiltonian's ground-state energy while assuming the molecule's nuclei remain fixed. Among the most studied algorithms for this problem are the Variational Quantum Eigensolver (VQE)~\cite{peruzzo:2014:VQE,kandala:2017:nature:VQE} for Noisy Intermediate-Scale Quantum (NISQ) devices and Quantum Phase Estimation (QPE)~\cite{Kitaev:1995:QuantumMA,nielsen:2010:quantum} for fault-tolerant quantum computers.
While these approaches show promise,  they face practical limitations. For instance, estimating the ground-state energy of FeMoCo, a well-known and practical benchmark in quantum chemistry, would require a fault-tolerant quantum computer with millions of physical qubits operating for nearly four days using QPE~\cite{femoco:2021:prxquantum}. In contrast, VQE can utilize fewer, noisier qubits, but its scalability is hindered by the extensive measurement demands during optimization, particularly for large-scale molecular systems~\cite{jules:2022:physrep:vqe:review,gonthier:2022:VQEblock}.

In practical quantum chemistry, estimating the ground-state energy for a single molecular configuration is often insufficient. Determining dynamic and structural properties, such as reaction barriers and optimal geometries, necessitates exploring multiple configurations. This requires knowledge of a family of ground states for a series of Hamiltonians parameterized by variables like nuclear coordinates and electron-nucleus distances. Consequently, one must compute numerous ground states with corresponding energies over a potential energy surface. However, a straightforward approach, such as independently running QPE or VQE for each configuration, incurs prohibitive computational costs. 
To address this, meta-based approaches leverage classical ML to optimize circuit parameters across multiple configurations simultaneously, as seen in Meta-VQE~\cite{cervera:2021:metaVQE}. Another intriguing approach is the use of generative QML to produce ground states with PQCs learning from quantum data~\cite{ceroni2023:gen-gr}. Nevertheless, these methods lack adaptability to varying molecule types and Hamiltonian forms and fail to capture correlations as effectively as Transformer-based models.

In this paper, we introduce the Molecular Quantum Transformer (MQT), a Quantum Transformer model designed to calculate molecular ground-state energies [Fig.~\ref{Figure:overview}(a)].
The MQT leverages attention mechanisms implemented through quantum circuits, enabling efficient modeling of complex interactions and correlations within molecular quantum systems. By training on random molecular configurations at each iteration, the MQT captures these interactions, allowing it to generalize and obtain ground-state energies across diverse configurations.
The MQT enables the simultaneous learning of ground-state energies for molecules across a range of bond lengths within a single model, providing greater resource efficiency compared to independently executing QPE or VQE for each molecular configuration [Fig.~\ref{Figure:overview}(b)].
We also compare MQT to a classical Transformer model under identical conditions of model dimensionality, demonstrating the superior performance of MQT over its classical counterpart.
Furthermore, the MQT can be utilized to learn new molecules efficiently through pretraining with diverse molecular data. This capability is particularly significant, as it allows for the applications of MQT in complex molecules with minimal effort, leveraging well-known molecular data.

\section*{Results}
\vspace{-0.5cm}

\subsection*{The electronic structure problem}
\vspace{-0.42cm}
\noindent
We consider a complex quantum many-body system in which multiple electrons and nuclei interact with each other through Coulomb interactions. Since the nuclei are thousands of times heavier than the electrons, they hardly move under the attraction from the electrons and can be regarded as fixed at coordinates $\bR_m$ (Born-Oppenheimer approximation). The wave function where $N$ electrons move around $M$ (fixed) nuclei (with atomic numbers $Z_1, \ldots, Z_m$) can be written as $\psi(\br_1, \br_2, \ldots, \br_N)$, and the Hamiltonian $H$ for the electrons can be expressed in Hartree atomic units as follows in the following simplified form (first quantization):
\begin{align}\label{eqn:elec:Hamiltonian}
    H(\bR) = -\sum_{i=1}^{N} \frac{1}{2} \nabla_i^2 + \sum_{i<j}^{N} \frac{1}{|\br_i - \br_j|} - \sum_{i=1}^{N} \sum_{m=1}^{M} \frac{Z_m}{|\br_i - \bR_m|}. 
\end{align}
Solving the Schrödinger equation $H\psi = E\psi$ with the energy eigenvalue $E$ of $H$ to determine the electronic state of a quantum many-body system is known as the molecular electronic problem. This problem is crucial for materials design and drug discovery but remains a central challenge in quantum chemistry due to the exponential complexity of quantum many-body systems.

\begin{figure*}
	\includegraphics[width=19cm]{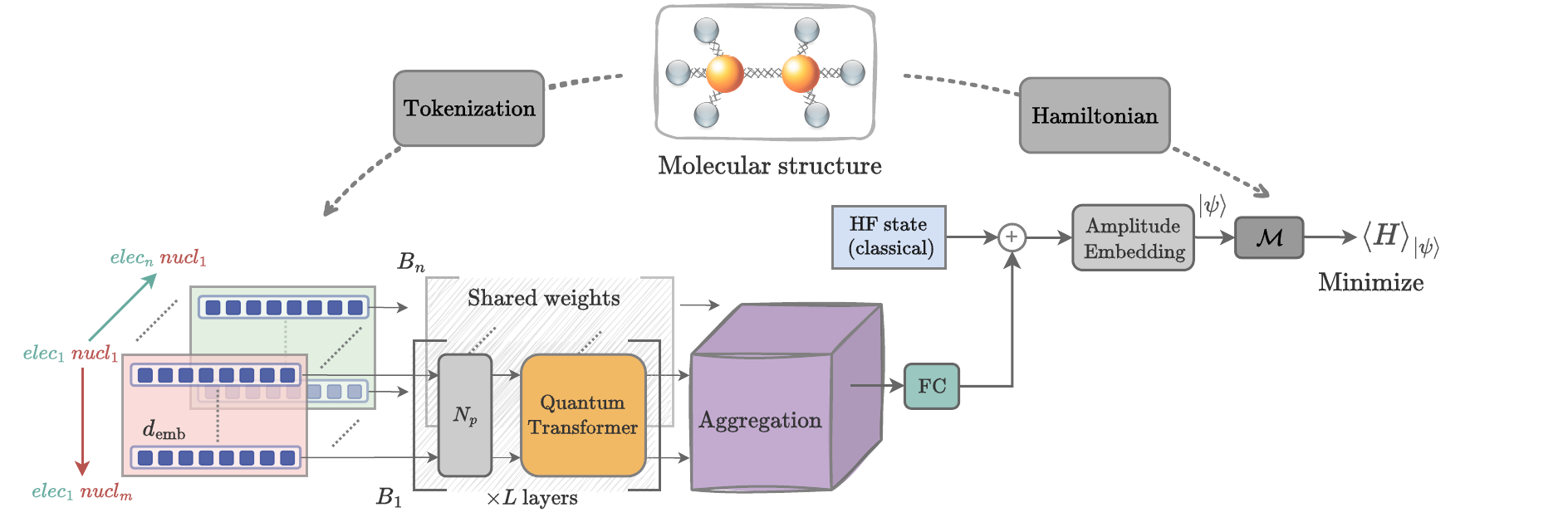}
	\protect{\caption{Structure of Molecular Quantum Transformer (MQT). Starting with a molecule defined by atomic symbols and nuclear coordinates, the molecular Hamiltonian $H$ is constructed in a qubit-based representation using $n_q$ qubits. The MQT tokenizes the electronic state into an $n\times m \times \demb$ feature matrix, where $\demb$ is the embedding dimension, $n$ and $m$ represent the number of electrons and nuclei, respectively. Tokens $elec_{i}\mathchar`-nucl_{j}$ ($i=1,\ldots,n$; $j=1,\ldots,m$) are processed by blocks $B_i$. Each block $B_i$ comprises $L$ layers, including an amplification module $N_p$ that scales features by proton number $N_p$ and a Quantum Transformer module with shared trainable parameters. Outputs are aggregated, mapped via an FC module to match the $n_q$-qubit state vector, and combined with the Hartree-Fock (HF) state to produce the final state through amplitude embedding. The expectation of the Hamiltonian measured on this final state is then minimized by optimizing the model's parameters.}\label{Figure:MQT}}
\end{figure*}

Classical methods such as Configuration Interaction (CI)~\cite{Szabo1996}, Coupled Cluster theory (CC)~\cite{Bartlett2007}, Møller–Plesset perturbation (MP)~\cite{moller:1934}, Density Functional Theory (DFT)~\cite{Kohn1996}, Quantum Monte Carlo (QMC)~\cite{Hammond1994}, and density matrix renormalization group (DMRG)~\cite{white:1992:dmrg} have significantly advanced the approximation of molecular electronic structure, each balancing accuracy and computational cost in different ways. In parallel, quantum computing offers a promising alternative by natively handling quantum states, with algorithms like QPE~\cite{Kitaev:1995:QuantumMA,nielsen:2010:quantum} and VQE~\cite{peruzzo:2014:VQE,kandala:2017:nature:VQE} emerging as leading approaches. While QPE provides exponential precision, it requires fault-tolerant hardware. In contrast, VQE is more practical for near-term devices but relies heavily on ansatz design and optimization strategies. We refer to the Methods section for more detailed explanations.

To be implemented on quantum computers, the Hamiltonian form in Eq.~\eqref{eqn:elec:Hamiltonian} is transformed into the following form (the second quantization): 
\begin{align}\label{eqn:hamiltonian:second}
    H(\bR) = \sum_{p,q}h_{pq}(\bR)c_p^{\dagger}c_q + \dfrac{1}{2}\sum_{p,q,r,s}h_{pqrs}(\bR)c_p^{\dagger}c_q^{\dagger}c_rc_s.
\end{align}
This operation is equivalent to the basis function expansion of the wave function. Here, $c_p^{\dagger}$ and $c_p$ are the fermionic creation and annihilation operators acting on the $p$-th orbital. Therefore, $c_p^{\dagger}c_q$ is the operator to transit the state on the $q$-th orbital to the $p$-th orbital, and $c_p^{\dagger}c_q^{\dagger}c_rc_s$ is the operator to transit a pair of state $(r,s)\to (p,q)$. The one and two-electron integrals $h_{pq}(\bR)$ and $h_{pqrs}(\bR)$ in the molecular orbital basis $\phi_p(\br)$ (yielded from the Hartree-Fock optimization procedure) depend implicitly on $\bR$, which is the general nuclear coordinate associated with the fixed nuclear configuration in 3-dimensional space.
These integral coefficients are calculated on the classical computer as follows:
\begin{align}
    h_{pq}(\bR) &= \int d\br \, \phi_p^*(\br) \left( -\frac{\nabla^2}{2} - \sum_m \frac{Z_m}{|\br - \bR_m|} \right) \phi_q(\br),\\
    h_{pqrs}(\bR) &= \int d\br_1 d\br_2 \, \frac{\phi_p^*(\br_1) \phi_q^*(\br_2) \phi_r(\br_2) \phi_s(\br_1)}{|\br_1 - \br_2|}.
\end{align}
The next step in implementing the Hamiltonian form in Eq.~\eqref{eqn:hamiltonian:second} using quantum circuits is to map the fermionic creation and annihilation operators to Pauli operators using methods such as the Jordan-Wigner or Bravyi-Kitaev transforms~\cite{seeley:2012:transform}.

\subsection*{Classical Transformer}
\vspace{-0.42cm}
\noindent
The Transformer architecture~\cite{attention:2017:NIPS} has become a crucial and widely adopted deep learning model in modern artificial intelligence. Originally developed to improve natural language processing capabilities, Transformers overcome the limitations of earlier deep learning models by effectively representing long-range dependencies and capturing complex relationships within data. The core innovation is the self-attention mechanism, which simultaneously captures correlations among all elements in a sequence, in contrast to the incremental processing of traditional models like recurrent neural networks. This parallel processing capability substantially decreases training time while also improving learning performance.

The standard architecture of the Transformer is illustrated in Fig.~\ref{Figure:Transformer} (see Methods) with both encoder and decoder components.  The encoder processes the source sequence through multiple layers, each combining multi-head self-attention with position-wise feed-forward networks. These layers are further enhanced with positional encodings, layer normalization, and residual connections to effectively capture sequential dependencies. The decoder employs masked multi-head self-attention to process the target sequence while simultaneously leveraging multi-head cross-attention to incorporate context from the encoder. The output of the decoder is refined by an additional feed-forward network and a linear layer, which generates the predictions.

In modern large language models (LLMs), the decoder-only architecture is primarily adopted. This design predicts the next token based solely on previous tokens, making it ideal for generating coherent, context-aware text. The streamlined decoder-only structure simplifies training and scaling, enabling efficient processing of vast datasets and massive model sizes across various natural language processing tasks.
Since our MQT is not developed for autoregressive language modeling, we only rely on the encoder-only Transformer.
This encoder transforms the sequence of positions of molecules in different configurations into a set of contextualized representations before forwarding to the energy minimization problem. 
We refer to the Methods section for the detailed steps of the self-attention mechanism and the encoder layer.

\subsection*{Molecular Quantum Transformer}
\vspace{-0.42cm}
\noindent
We propose the Molecular Quantum Transformer (MQT) model, which replaces the VQE ansatz with a Quantum Transformer structure to determine the electron wave function that minimizes the energy associated with the Hamiltonian. The fundamental concept of the attention mechanism, described in the previous section, involves capturing correlations among all tokens simultaneously. In the context of the MQT, these tokens represent features derived from the positions and distances between atomic nuclei. As configurations change, the model adapts its features such as electron configurations based on these relationships. This adaptation is generalized through training the model across a variety of conditions. In contrast, running QPE or VQE demands independent resources for each molecular configuration, necessitating multiple models. The MQT, however, can learn the ground-state energies for various molecules and various configurations simultaneously within a single model.

The structure of our MQT model is illustrated in Fig.~\ref{Figure:MQT}. Given a molecular structure defined by atomic symbols and the nuclear coordinates of its constituent atoms, we construct the molecular Hamiltonian $H$ in a qubit-based representation using $n_q$ qubits as a preprocessing step. Here, we utilize the PennyLane molecules dataset~\cite{Utkarsh:2023:Chemistry} and its built-in functions~\cite{bergholm:2022:pennylane} to generate the corresponding molecular Hamiltonians.
The MQT begins with a tokenization module that creates input tokens $elec_{i}\mathchar`-nucl_{j}$ ($i=1,\ldots,n$; $j=1,\ldots,m$),  representing the electronic state as an $n\times m$ two-dimensional array of $\demb$-dimensional features. Here, $\demb$ is the embedding dimension, $n$ and $m$ denote the number of electrons and nuclei, respectively, resulting in an $n\times m \times \demb$ feature matrix.
For each electron index $i$, the $m$ tokens  $elec_{i}\mathchar`-nucl_{j}$ ($j=1,\ldots,m$) are processed by a block $B_i$, which consists of L layers. Each layer includes an amplification module $N_p$ and a Quantum Transformer module. 
The amplification module $N_p$ scales the feature values by the proton number $N_p$ to reflect the differences among nuclear species before passing them to the Quantum Transformer module.
The Quantum Transformer module contains trainable parameters, which are shared across all blocks $B_1,\ldots, B_n$.
The outputs of all Quantum Transformer modules are combined into a single feature representation via an aggregation module. A fully connected (FC) module then maps this aggregated feature into a vector with the same dimensionality as the $n_q$-qubit state vector. This transformed vector is subsequently added to the Hartree-Fock (HF) state vector, and the resulting representation is used to generate the final electronic state through an amplitude embedding module. The expectation value of the Hamiltonian $H$ is computed from this amplitude-embedded state via a measurement process. This expectation value is minimized by optimizing the trainable parameters in the Quantum Transformer, aggregation, and FC modules.
We refer to the Methods section for the detailed structures of these modules.

\begin{figure*}
	\includegraphics[width=16cm]{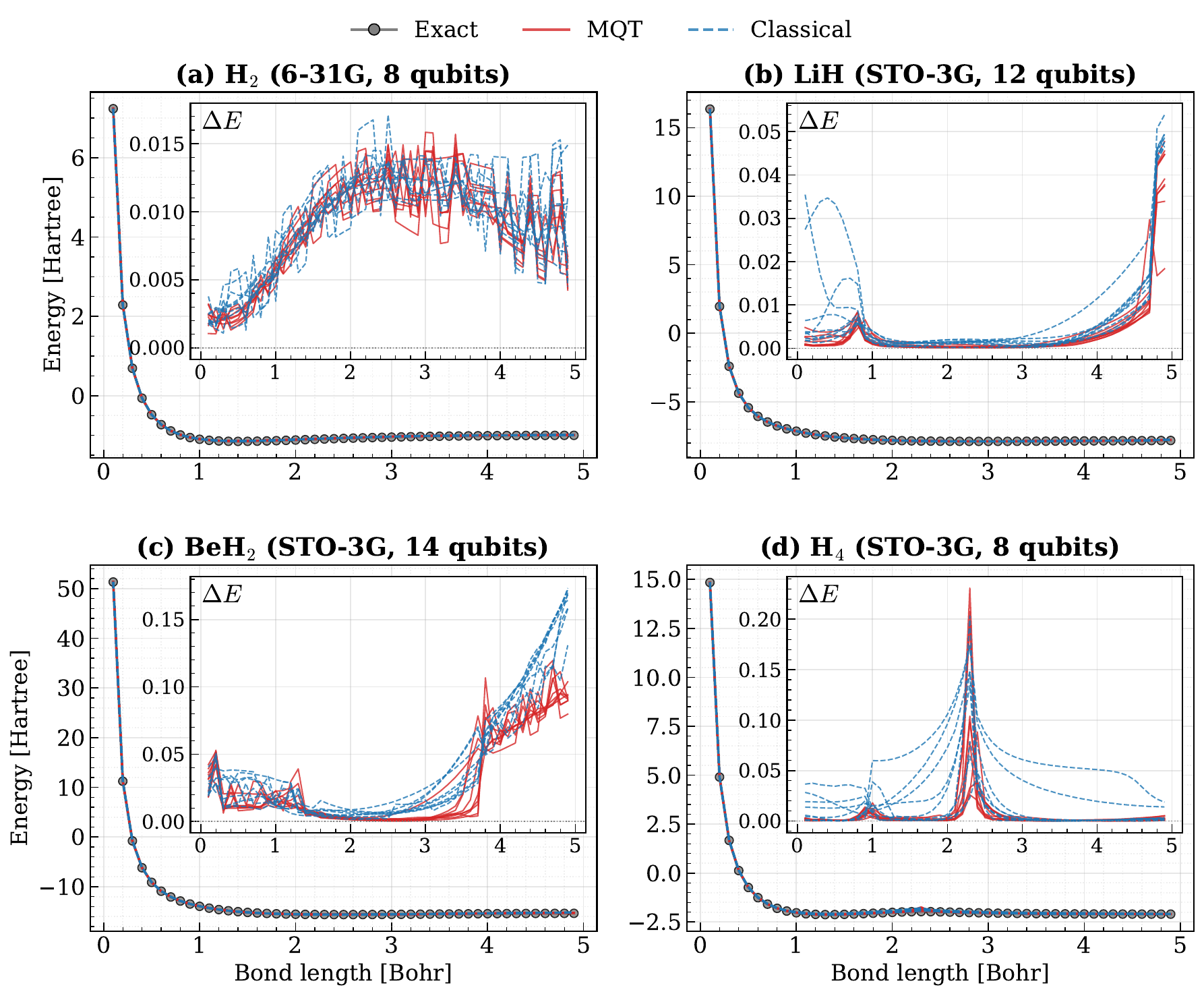}
	\protect{\caption{Potential energy curves and estimation errors ($\Delta E$) in the inset plots for varying interatomic bond lengths in (a) $\textup{H}_{2}$ (b) LiH, (c) $\textup{BeH}_{2}$, and (d) $\textup{H}_{4}$ molecules using the quantum (red lines) and classical (dotted blue lines) Transformers. In the main plots, the averages of the exact results (gray line with circle markers), MQT, and classical methods over nine trials are shown, but they overlap within the displayed range.\label{Figure:EnegyCurve}}}
\end{figure*}

\begin{figure}
	\includegraphics[width=8.7cm]{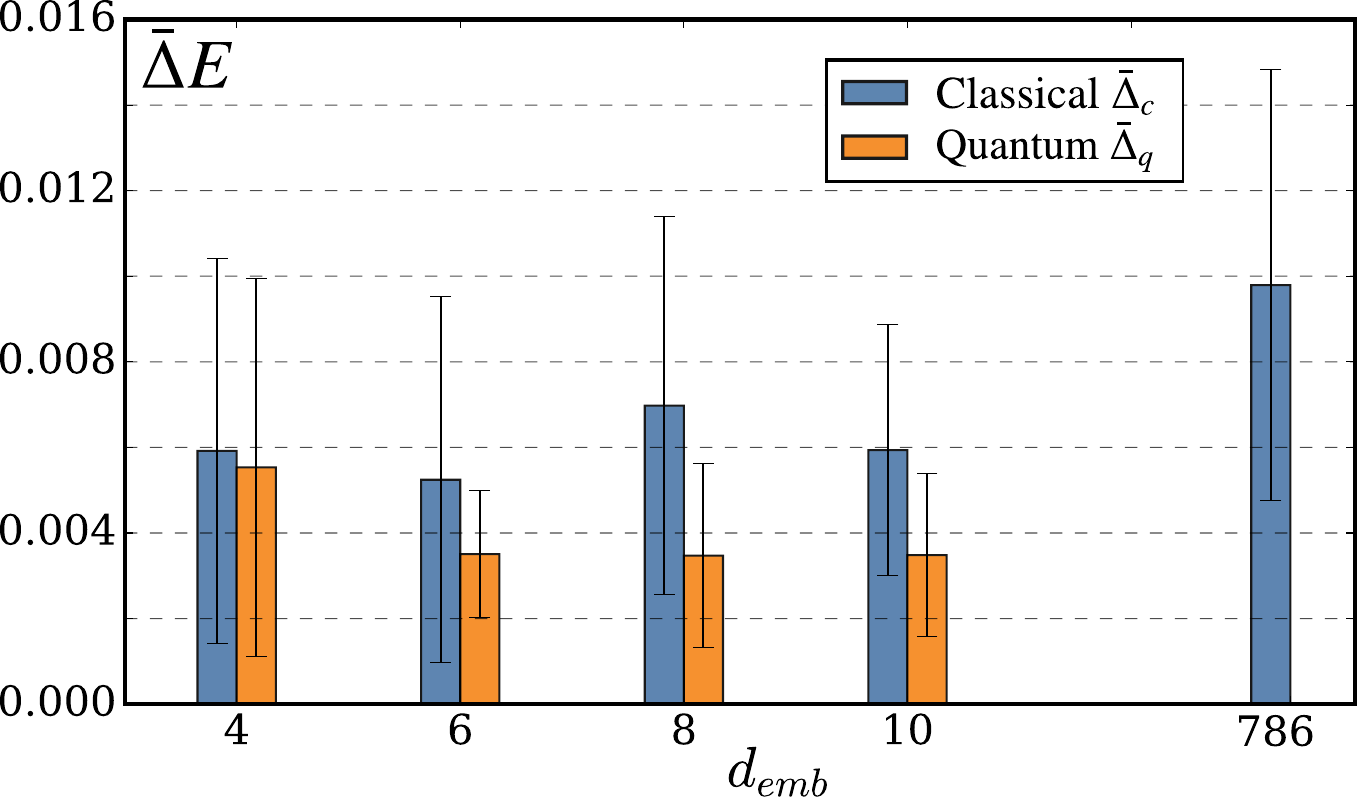}
	\protect{\caption{Bar plot comparing the average ground-state energy estimation error of LiH over potential energy curves between the classical Transformer ($\overline{\Delta}_c$) and MQT ($\overline{\Delta}_q$) as a function of the token embedding dimension $\demb$ (4, 6, 8, 10, and 786). Error bars represent standard deviations. Teal-blue bars indicate classical results, while orange bars indicate quantum results, highlighting MQT’s consistently lower errors across $\demb$. The quantum result at $\demb = 786$ is not displayed due to computational resource limitations.
\label{Figure:tokendim}}}
\end{figure}

\subsection*{Numerical experiments}
\vspace{-0.42cm}
\noindent
In the following numerical experiments, we apply MQT to estimate the potential energy curves (PEC) of the molecular Hamiltonian for $\textup{H}_{2}$, $\textup{LiH}$, $\textup{BeH}_{2}$, and $\textup{H}_{4}$.
For the second quantization of the Hamiltonians, we employ the Bravyi-Kitaev mapping for $\textup{H}_{2}$ (8 qubits), $\textup{LiH}$ (12 qubits), and $\textup{BeH}_{2}$ (14 qubits), and the Jordan-Wigner transformation for $\textup{H}_{4}$ (8 qubits). The 6-31G basis set is used for $\textup{H}_2$, while STO-3G basis set is applied to the other molecules.

We first evaluate MQT in a plain training scenario, where it is trained and tested on data from the same molecule. Subsequently, we investigate the pretraining effect by training MQT on $\textup{H}_{2}, \textup{BeH}_{2}$, and  $\textup{H}_{4}$, followed by fine-tuning on a different molecule ($\textup{LiH}$).
We set the embedding dimension $\demb=8$ as the default setting for these experiments.

\subsubsection*{Estimating the potential energy curve with plain training}

\begin{figure}
	\includegraphics[width=8.7cm]{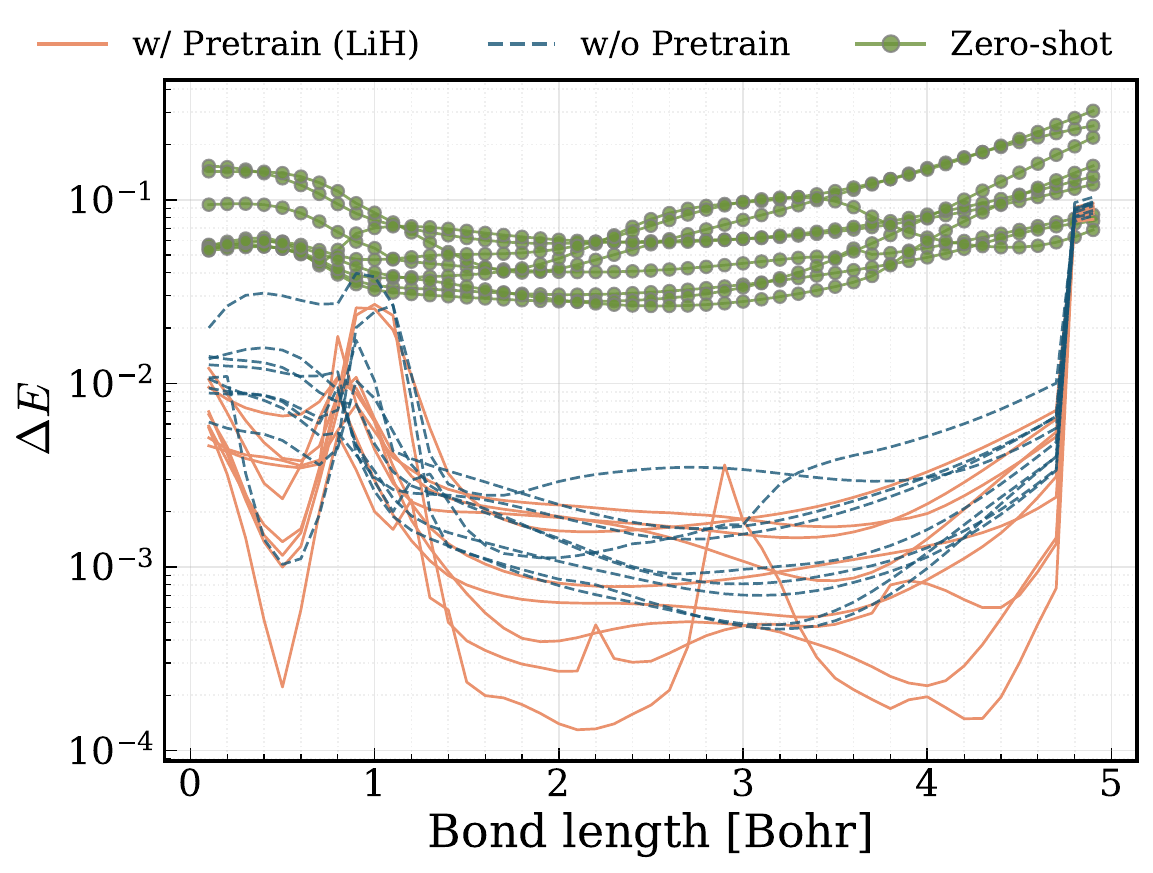}
	\protect{\caption{Energy estimation errors ($\Delta E$) in nine trials show MQT trained on LiH with a few data points (few-shot learning), comparing pretraining on $\textup{H}_{2}$, $\textup{BeH}_{2}$, and $\textup{H}_{4}$ (orange lines)  versus no pretraining (dotted teal-blue lines) and zero-shot learning (green line with circle markers) across varying bond lengths.
    Here, zero-shot MQT is pretrained on $\textup{H}_{2}$, $\textup{BeH}_{2}$, and $\textup{H}_{4}$ but not fine-tuned on LiH.
    The pretrained MQT provides more accurate estimations in few-shot learning than MQT models relying solely on zero-shot learning or fine-tuning. It even outperforms the neural network-based meta-VQE~\cite{cervera:2021:metaVQE} when trained from scratch, showing reductions in average error by approximately 19\% and 11\%, respectively. 
    \label{Figure:fewshot}}}
\end{figure}

In plain training, the token matrix is generated in each iteration from a configuration with a random bond length sampled from  (0.0, 5.0) [Bohr]. After training, the potential energy curves are estimated for bond lengths ranging from 0.1 to 4.9 [Bohr] in 0.1 [Bohr] increments using the trained model.
We use $L=6$ layers of Quantum Transformer module in each processing block $B_i$ as shown in Fig.~\ref{Figure:MQT}.
We compare the performance of MQT against a classical Transformer implementation replacing the Quantum Transformer module in Fig.~\ref{Figure:MQT}.
To align with the settings of MQT, we consider the classical Transformer with $\demb$ embedding dimensions, $4\times\demb$ hidden dimensions (see Methods for the definition of the hidden dimension), and a single-head attention.
We use AdamW~\cite{loshchilov:2018:decoupled} optimizer with a weight decay rate of 0.001. The learning rate is set to 0.008 for the classical Transformer with $\textup{H}_{2}$; and 0.004 for other cases. In the optimization, we employ the method in Ref.~\cite{hogwild:2011:NIPS} to run four processes in parallel on a single GPU, with each process performing 2500 iterations, for a total of $10^4$ iterations. We use the TorchQuantum~\cite{hanruiwang:2022:quantumnas} library for the simulation of quantum circuits in MQT.

\begin{table}[h!]
    \centering
    \resizebox{\columnwidth}{!}{ 
    \begin{tabular}{c c c c c}
        \hline
        \hline
        Metric & $\textup{H}_2$ & LiH & $\textup{BeH}_2$ & $\textup{H}_4$ \\
        \hline
        $\overline{\Delta}_c$ & 9.3\textup{e-3}\qquad & 6.0\textup{e-3} & 3.3\textup{e-2} & 1.9\textup{e-2}\\
        $\overline{\Delta}_q$ & 8.9\textup{e-3}\qquad & 3.5\textup{e-3} & 2.7\textup{e-2} & 0.5\textup{e-2}\\
        \hline
        \hline
    \end{tabular}
    }
    \caption{Average ground-state energy estimation error by classical Transformer ($\overline{\Delta}_c$) and MQT ($\overline{\Delta}_q$)}
    \label{tab:res:deltaE}
\end{table}

Figure~\ref{Figure:EnegyCurve} compares the energy estimations of the MQT and classical Transformer against the theoretically calculated ground-state energies of the Hamiltonians for each molecule. The inset figures show the estimation errors relative to the true ground-state energy, with each line representing one of nine trials. MQT exhibits lower estimation errors than the classical Transformer for (b) LiH, (c) $\textup{BeH}_{2}$, and (d) $\textup{H}_{4}$. 
For (a) $\textup{H}_{2}$, the estimation errors of MQT are nearly identical to those of the classical Transformer (note the differing y-axis scales across plots). The average estimation errors across all tested bond lengths for the classical Transformer ($\bar{\Delta}_c$) and MQT ($\bar{\Delta}_q$) are summarized in Tab.~\ref{tab:res:deltaE}.
These results suggest that MQT performs comparably to the classical Transformer for $\textup{H}_{2}$ but outperforms it for other molecules, reducing the average estimation error by 42\% in LiH, 18\% in $\textup{BeH}_{2}$, and 74\% in $\textup{H}_{4}$ compared to the classical Transformer. 
In Fig.~\ref{Figure:tokendim}, we further compare $\overline{\Delta}_c$ and $\overline{\Delta}_q$ of LiH for different embedding dimension $\demb \in \{4, 6, 8, 10\}$. MQT shows lower estimation error consistency with all $\demb$ and achieves the saturated value when $\demb \geq 6$.
Interestingly, even with $\demb = 786$, the classical Transformer exhibits significantly higher estimation errors compared to MQT at much lower $\demb$. This discrepancy may be attributed to the optimization difficulty associated with the large number of parameters in the classical model, or it may indicate that MQT is inherently better suited to modeling molecular data. These findings underscore the potential advantages of integrating quantum structures into the Transformer framework.

\subsubsection*{Estimating the potential energy curve with pretraining}

The MQT model, pretrained on multiple molecules ($\textup{H}_{2}$, $\textup{BeH}_{2}$, $\textup{H}_{4}$) is fine-tuned with a small number of data points for $\textup{LiH}$. Here, a few-shot learning approach is performed to estimate the potential energy curve of $\textup{LiH}$.

During fine-tuning, the training mirrors the setting in the pretraining stage, employing the AdamW optimizer with a weight decay of $10^{-3}$ and a learning rate of $4\times 10^{-3}$. The training data for LiH consisted of five randomly selected bond lengths: $\{0.5, 1.5, 2.5, 3.5, 4.5\}$ [Bohr]. 
The method in Ref.~\cite{hogwild:2011:NIPS} is employed to run four processes in parallel on a single GPU, with each process performing 500 iterations, for a total of 2000 iterations for fine-tuning.
After the fine-tuning, potential energy curves are estimated for bond lengths from 0.1 to 4.9 Bohr in 0.1 Bohr increments.
Figure~\ref{Figure:fewshot} depicts the average estimated error curves for pretraining and fine-tuning (orange lines, labeled  ``w/ Pretrain (LiH)") and fine-tuning only (green lines, labeled ``w/o Pretrain") across nine trials.
The pretrained model yields more accurate estimations in few-shot learning compared to only using fine-tuning, as indicated by reducing nearly 19\% of error to the theoretical values from $7.6\times 10^{-3}$ (``w/o Pretrain") to $6.2\times 10^{-3}$ (``w/ Pretrain (LiH)"). 
Notably, with the same number of training data points, our model achieves an 11\% improvement in few-shot learning performance compared to the neural network-based meta-VQE~\cite{cervera:2021:metaVQE}, which yields an average estimation error of $7.0 \times 10^{-3}$. The neural network-based meta-VQE, identified as the most effective method in Ref.~\cite{cervera:2021:metaVQE}, was reimplemented in our experiments on LiH.

To further examine the effect of few-shot learning, we evaluate the zero-shot learning, where the pre-trained MQT is tested on LiH without fine-tuning. As shown in Fig.~\ref{Figure:fewshot}, the estimation errors of zero-shot MQT significantly exceed those of few-shot MQT, indicating that zero-shot learning is inadequate. At least a few data points (few-shot learning) are necessary during fine-tuning to enhance estimation accuracy.

\section*{Discussion}
\vspace{-0.5cm}
\noindent
We proposed the MQT model, which leverages the quantum attention mechanism to revolutionize the calculation of ground-state energies.
The key contribution is that MQT can be adapted to train on multiple molecules and multiple configurations without altering the model structure. 
This approach can open new avenues in quantum chemistry and materials science, where an accurate understanding of ground-state energies is crucial, but running routines such as VQE and QPE independently for each molecule and configuration requires high cost.

In our numerical experiment, we evaluated the performance of MQT in estimating potential energy curves for $\textup{H}_{2}$, $\textup{LiH}$, $\textup{BeH}_{2}$, and $\textup{H}_{4}$, benchmarking it against a classical Transformer. 
MQT consistently outperformed the classical model in estimation accuracy.
We also investigated pretraining MQT on multiple molecules ($\textup{H}_{2}$, $\textup{BeH}_{2}$, and $\textup{H}_{4}$)
and evaluated the few-shot learning capability on LiH.
Results showed that pretraining modestly improved accuracy for the LiH potential energy curve compared to an untrained model. In scenarios with small molecules and readily available training data, pretraining MQT can significantly reduce the number of quantum circuit runs required for larger and more complex molecular systems. To maximize this pretraining advantage, designing suitable pretraining datasets for specific molecules is crucial. Ideally, with sufficiently high-quality pretraining data, MQT could estimate the potential energy surface of a given molecule without a few-shot learning on that molecule.

Scaling MQT to larger molecules requires addressing several limitations. A primary bottleneck is the output quantum state representation via amplitude embedding, where direct state preparation is challenging due to the complexity scaling exponentially with the number of qubits $n_q$. The qRAM~\cite{lloyd:2008:QRAM} scheme could reduce this complexity to near-linear in $n_q$, but fault-tolerant qRAM remains an engineering challenge. Even with efficient qRAM, preparing the state vector on a classical computer prior to amplitude embedding demands memory that grows exponentially with $n_q$, hindering scalability. 
Inspired by the divide-and-conquer approach in VQE~\cite{fujii:2022:VQE:divide}, we can concatenate MQTs to enable the application to large systems with strong intrasubsystem and weak inter-subsystem interactions on small-scale quantum computers. Thus, our method still offers a promising framework for tackling practically significant, large-scale problems in quantum chemistry.

In this paper, MQT is developed to exploit the classical representation of a molecule and its associated Hamiltonian, enabling efficient computation of electronic properties within a quantum framework. This approach necessitates an embedding procedure to transform the classical representation into quantum features suitable for processing on quantum hardware. As a future direction, MQT could be extended to integrate directly with quantum-native representations of molecular states, bypassing classical intermediaries. For instance, incorporating the noisy ground state estimates obtained from VQE or the intermediate eigenstate data from the QPE process could enrich MQT with detailed quantum information about the ground state and low-lying excited states. Such an integration promises to enhance the accuracy of MQT by leveraging the intrinsic quantum correlations captured in these states, which are often inaccessible to classical methods like DFT or CI. Furthermore, this approach could be combined with advanced quantum techniques, such as quantum denoising mechanisms~\cite{tran:2024:varQAE} to mitigate errors from NISQ devices, or quantum curriculum learning algorithms~\cite{tran:2024:qcurl} to optimize the training of quantum circuits. These enhancements could position MQT as a versatile tool for tackling the molecular electronic problem, potentially outperforming traditional hybrid quantum-classical workflows in both precision and scalability.

In the near term, while improvements in the architectural completeness of Quantum Transformers and their compatibility with classical components are still necessary, establishing a unified benchmarking framework is critical for accurately evaluating their real-world performance across different approaches. Generative tasks involving molecular data represent particularly promising applications in this context.
Looking ahead to the early era of fault-tolerant quantum computing, Quantum Transformer models based on quantum linear algebra offer strong theoretical potential~\cite{guo2024:qla:transformer,liao:2024:gptquantumcomputer,khatri:2024:quixer:transformer}. 
As quantum hardware continues to evolve, Quantum Transformers are expected to play an increasingly impactful role in solving complex tasks, unlocking new possibilities in quantum-enhanced machine learning. In this perspective, MQT serves as a compelling example of how quantum-enhanced architectures can go beyond traditional text and image applications, addressing fundamental challenges in quantum chemistry and materials science, domains where classical neural networks are inherently limited.

\section*{Methods}
\vspace{-0.5cm}

\subsection*{The electronic structure problem}\label{appx:elec}
\vspace{-0.42cm}
\noindent
Advanced methods in classical computation have advanced our ability to approximate solutions to the molecular electronic problem. These include the configuration interaction (CI) method, which constructs wavefunctions as linear combinations of Slater determinants to capture electron correlation~\cite{Szabo1996}; the coupled cluster (CC) expansion~\cite{Bartlett2007}, which employs an exponential ansatz to account for electron correlation more efficiently than CI; and Møller–Plesset perturbation theory (MP)~\cite{moller:1934}, which treats electron correlation as a perturbative correction to the Hartree-Fock solution. Density functional theory (DFT)~\cite{Kohn1996}, another cornerstone method, focuses on electron density rather than the wave function, offering a balance between accuracy and computational cost. The quantum Monte Carlo (QMC)~\cite{Hammond1994} method provides a stochastic approach to sampling the wavefunction, achieving high accuracy for small systems despite its computational intensity. The density matrix renormalization group (DMRG) method~\cite{white:1992:dmrg}, originally developed for condensed matter physics, has been adapted to quantum chemistry and is particularly effective for systems with strong electron correlation~\cite{Chan2011}. 


In parallel, quantum computers have emerged as a promising alternative platform for addressing the molecular electronic problem, leveraging their ability to naturally represent and manipulate quantum states. Quantum computers offer the potential to overcome the limitations of classical algorithms, particularly for preparing ground states of quantum many-body systems that are intractable with traditional methods. Significant efforts have focused on developing quantum algorithms tailored to this task. Quantum Phase Estimation (QPE)~\cite{Kitaev:1995:QuantumMA,nielsen:2010:quantum} extracts eigenvalues of a Hamiltonian with exponential precision, provided a good initial state is available. Variational Quantum Eigensolver (VQE)~\cite{peruzzo:2014:VQE,kandala:2017:nature:VQE} adopts a hybrid quantum-classical approach to variationally minimize the energy expectation value. Other approaches include adiabatic quantum computing, which evolves a system from a simple initial Hamiltonian to the target Hamiltonian~\cite{farhi:2000:adiabatic}; imaginary-time evolution, which simulates non-unitary dynamics to converge to the ground state~\cite{McArdle:2019:imagine}; and subspace~\cite{motta:2023:subspace} and Lanczos~\cite{Kirby:2023:lanczos} methods, which reduce the problem's dimensionality by focusing on a subset of the Hilbert space. 

Among these quantum algorithms, QPE and VQE stand out as flagship algorithms due to their theoretical rigor and experimental feasibility, respectively.
QPE is theoretically exact but requires deep circuits and large qubit counts, making it best suited for future error-corrected quantum computers. VQE, on the other hand, combines shallow quantum circuits with classical optimization to minimize the energy expectation value of a parameterized quantum state (ansatz). VQE is more feasible on NISQ devices, though its accuracy depends on the choice of ansatz and optimization strategy~\cite{jules:2022:physrep:vqe:review}.

\subsection*{Classical Transformer}\label{appx:transformer}
\begin{figure}
	\includegraphics[width=8.7cm]{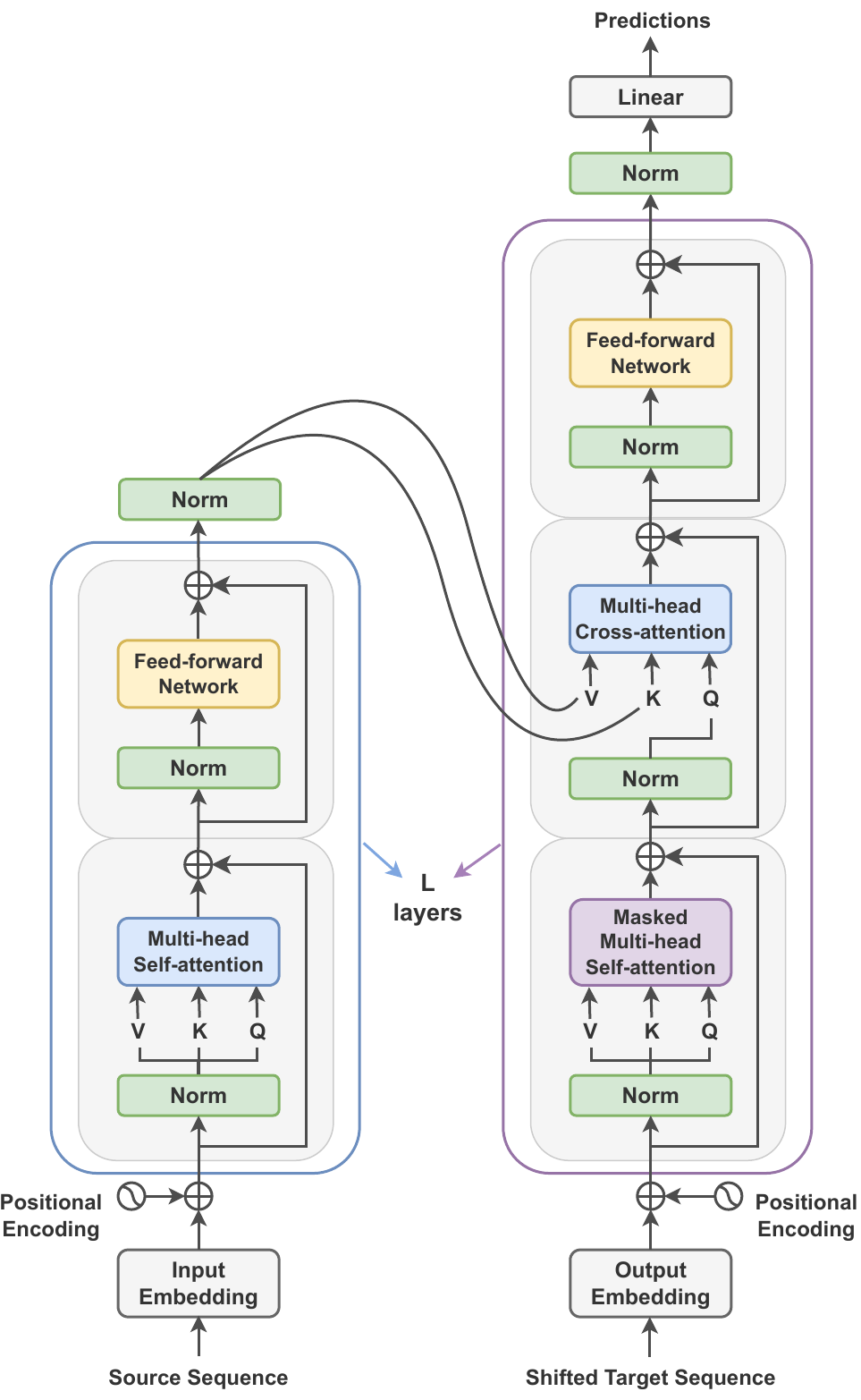}
	\protect{\caption{Standard architecture of classical Transformer with an encoder (left) and a decoder (right). 
    The encoder processes the source sequence through layers that combine multi-head self-attention and position-wise feed-forward networks, enhanced with positional encodings, layer normalization, and residual (bypass) connections. The decoder applies masked multi-head self-attention to the target sequence and cross-attention to integrate encoder context, with its output refined by a feed-forward network and a linear layer to generate predictions.
    The figure has been recreated to clearly detail the sequential operations originally shown in Ref.~\cite{attention:2017:NIPS} and modified in Ref.~\cite{godoy:2020:dlvisuals}.\label{Figure:Transformer}}}
\end{figure}

\vspace{-0.42cm}
\noindent
The standard architecture of the Transformer is illustrated in Fig.~\ref{Figure:Transformer} with both the encoder and decoder components.  The encoder processes the source sequence through multiple layers, each combining multi-head self-attention with position-wise feed-forward networks. These layers are further enhanced with positional encodings, layer normalization, and residual connections (the arrows bypassing the main components, such as the attention and feed-forward layers) and to effectively capture sequential dependencies. The decoder employs masked multi-head self-attention to process the target sequence while simultaneously leveraging multi-head cross-attention to incorporate context from the encoder. The output of the decoder is refined by an additional feed-forward network and a linear layer, which generates the predictions.
In our MQT, we only rely on the encoder-only Transformer to transform the position sequence of molecules in different configurations into a set of rich and contextualized representations. Therefore, in the remainder of this section, we focus on detailing the implementation of encoder-only Transformer architectures.

The encoder-only Transformer processes input sequences by first embedding the tokens with positional encodings, then passing them through multiple identical layers. Each layer applies multi-head self-attention to capture inter-token relationships, followed by a position-wise feed-forward network to refine the representations. Residual connections and layer normalization are employed throughout to ensure stable training and effective gradient flow. This architecture enables efficient, parallelizable processing while capturing complex dependencies within the data.

\subsubsection*{Tokenization and Input Representation}

Transformers employ tokenization to convert sequential data into discrete units. Tokenization breaks a sequence into smaller units called \textit{tokens}.
Given an input sequence of tokens \( x_1, x_2, \dots, x_n \), each token is first mapped into a continuous embedding vector. Let the embedding function be \( e(\cdot) \) so that $E=[e(x_1),e(x_2),\ldots,e(x_n)]^\top$,
where \( \demb \) is the embedding dimension.
Transformers do not inherently encode token order. Instead, a positional encoding \( P \in \mathbb{R}^{n \times \demb} \) is added to the embeddings as $Z^{(0)} = E + P$.

\subsubsection*{Self-attention and Encoder Layer}

The encoder is composed of \( L \) identical layers. Each layer has two main sub-layers: multi-head self-attention and a position-wise feed-forward network, with residual connections and layer normalization applied around each sub-layer.
Self-attention is the core component of the Transformer architecture with the capability to model token interdependencies. It allows each token in the sequence to dynamically incorporate information from all other tokens by computing inner-product-based attention scores. This operation, known as \textit{scaled dot-product attention}, produces attention matrices that quantify the relevance of each token relative to the others, effectively capturing complex contextual relationships across the entire sequence.
The Transformer also incorporates the multi-head attention mechanism, which extends self-attention by computing multiple attention matrices in parallel. Each attention head is trained to capture different relationships between tokens in the sequence, allowing the model to process information from multiple embedding subspaces simultaneously. Typically, Transformer uses high-dimensional embeddings, which are divided across several attention heads.

For the $l$-th layer $(l = 1, \dots, L)$ of the encoder, the input is $Z^{(l-1)}$ and the output is $Z^{(l)} = \textup{en}_l(Z^{(l-1)})$. The detailed computational steps of the $l$ layer are:

\begin{enumerate}
\item  \textbf{Layer normalization and projections:}
The input $Z^{(l-1)}$ is normalized across the features for each individual data point to $\hat{Z}^{(l-1)}$ and then linearly projected to form queries $Q$, keys $K$, and values $V$. For each attention head $i$ ($i= 1, \dots, h$), we have:
    \begin{align}
    Q_i = \hat{Z}^{(l-1)}W^Q_i,\quad K_i = \hat{Z}^{(l-1)}W^K_i,\quad V_i = \hat{Z}^{(l-1)}W^V_i,    
    \end{align}
    where trainable parameters $W^Q_i, W^K_i, W^V_i \in \mathbb{R}^{\demb \times d_k}$ and typically $d_k = \frac{\demb}{h}$.
    
\item \textbf{Scaled dot-product attention:}  
For each head, the attention output is computed as:
\begin{align}
\textup{head}_i = \textup{Attention}(Q_i, K_i, V_i) = \textup{softmax}\!\left(\frac{Q_i K_i^T}{\sqrt{d_k}}\right)V_i.
\end{align}
    
\item \textbf{Concatenation and projection:}  
The outputs of all heads are concatenated and projected:
\begin{align}
\textup{MultiHead}(Z^{(l-1)}) = \textup{Concat}\left(\textup{head}_1, \dots, \textup{head}_h\right)W^O,
\end{align}
where trainable parameters $W^O \in \mathbb{R}^{hd_k \times \demb}$.

\item \textbf{Residual connection and normalization:}  
The output of the multi-head attention is added to the original input \( Z^{(l-1)} \) (residual connection), and then normalized:
\begin{align}
\hat{Z}^{(l)} = \textup{LayerNorm}\Bigl(Z^{(l-1)} + \textup{MultiHead}(Z^{(l-1)})\Bigr).
\end{align}

\item \textbf{Position-wise feed-forward network (FFN): }
The normalized output $\hat{Z}^{(l)}$ is processed by a two-layer FFN applied independently to each position:
\begin{align}
\textup{FFN}(\hat{Z}^{(l)}) = \sigma\Bigl(\hat{Z}^{(l)}W_1 + b_1\Bigr)W_2 + b_2,
\end{align}
where trainable parameters $W_1 \in \mathbb{R}^{\demb \times d_{\textup{ff}}}$, $b_1 \in \mathbb{R}^{d_{\textup{ff}}}$, $W_2 \in \mathbb{R}^{d_{\textup{ff}} \times \demb}$, $b_2 \in \mathbb{R}^{\demb}$, and $\sigma(x)$ is an activation function such as $\textup{tanh}(x)$ and $\textup{ReLU}(x)=\textup{max}(0, x)$.
Here, $d_{\textup{ff}}$ is the hidden dimension of the FFN, typically larger than $\demb$

\item \textbf{Residual connection: }  
\begin{align}
Z^{(l)} = \textup{LayerNorm}\Bigl(\hat{Z}^{(l)} + \textup{FFN}(\hat{Z}^{(l)})\Bigr).
\end{align}

\end{enumerate}

The final encoder output through $L$ identical layers is:
\begin{align}
Z^{(L)} = \textup{LayerNorm}\Bigl(\textup{en}_L\Bigl(\dots \textup{en}_1\bigl(Z^{(0)}\bigr) \dots \Bigr)\Bigr).
\end{align}
This output \( Z^{(L)} \) is a set of context-rich representations for the input tokens, encapsulating both local and global dependencies, and can be used for various downstream tasks such as classification, regression, or serving as input to a decoder in sequence-to-sequence models.

\subsection*{Molecular Quantum Transformer}
\vspace{-0.42cm}
\noindent
We explain the detailed steps of the Tokenization procedure, the detailed structures of the Quantum Transformer and Aggregation modules in MQT.

\subsubsection*{Tokenization}

\begin{figure*}
	\includegraphics[width=16cm,page=1]{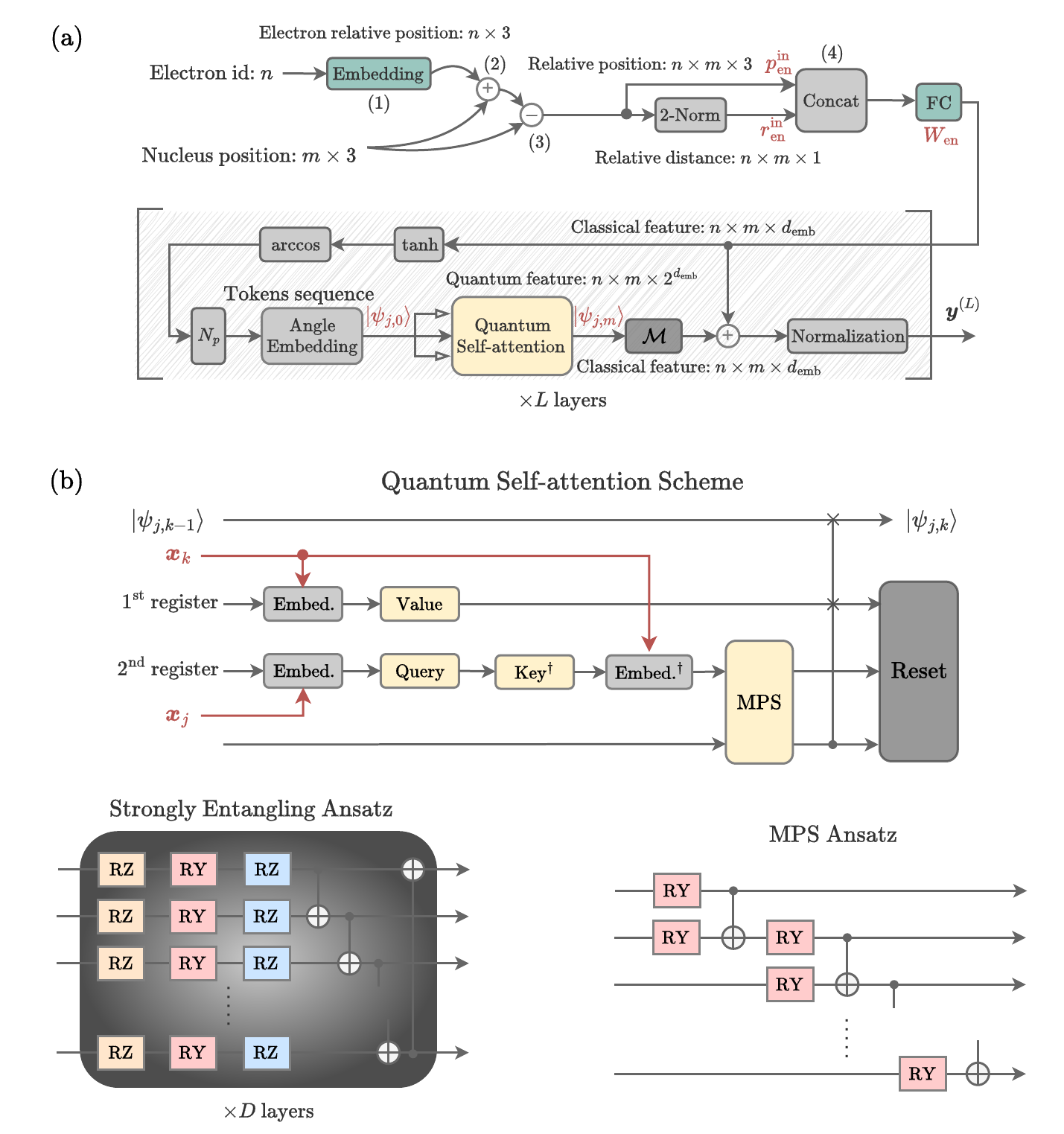}
	\protect{\caption{(a) Tokenization and Quantum Transformer module in the MQT. (Upper panel) Each token vector is derived from concatenated relative electron-nucleus positions ($p^{\textup{in}}_{\textup{en}}$) and distances ($r^{\textup{in}}_{\textup{en}}$), processed through a fully-connected (FC) layer with trainable weights $W_{\textup{en}}$. (Lower panel) The Quantum Transformer comprises $L$ repeated layers, processing input features with $\textup{arccos}$, $\textup{tanh}$, and amplification $N_p$ modules to generate an $m\times \demb$ features matrix of $m$ tokens $\bx_j$ for each electron. Each $\demb$-dimensional feature vector of token $\bx_j$ is embedded into a quantum state $\ket{\psi_{j,0}}$ via an angle embedding layer using $\demb$ qubits, updated via Query, Key, and Value transformations ($\ket{\psi_{j,k-1}}\to\ket{\psi_{j,k}}$ $(k=1,\ldots,m)$) using a quantum self-attention mechanism. The measurement in the final state $\ket{\psi_{j,m}}$ yields an $n\times m \times \demb$ feature matrix for $n$ blocks of $n$ electrons. These features are added to the input features via a residual connection and normalization to produce the input for the next layer.
    (b) Quantum self-attention update ($\ket{\psi_{j,k-1}}\to\ket{\psi_{j,k}}$). 
For token $\bx_j$, angle embedding with $\textup{R}_Y$ rotation gate ($\textup{Embed.}$) uses a second auxiliary token register, followed by Query and the adjoint Key transformations, which are constructed with a single layer of strongly entangling  (StrEnt) ansatz. Then the adjoint matrix of the angle embedding ($\textup{Embed.}^{\dagger}$) for the token $\bx_k$ is applied to compute the Hadamard product of the Query and Key. This product is then transformed into a 1-qubit attention representation using a 2-bond matrix product state (MPS) ansatz implemented with an additional ancilla qubit.
Value transformation for $\bx_k$ in the first auxiliary token register (a six-layer StrEnt ansatz after $\textup{Embed.}$) and a controlled SWAP gate complete the update. The StrEnt and MPS ansatzes are depicted in the lower panel with the trainable rotation gates.\label{Figure:attention}}}
\end{figure*}

\begin{figure*}
	\includegraphics[width=16cm,page=1]{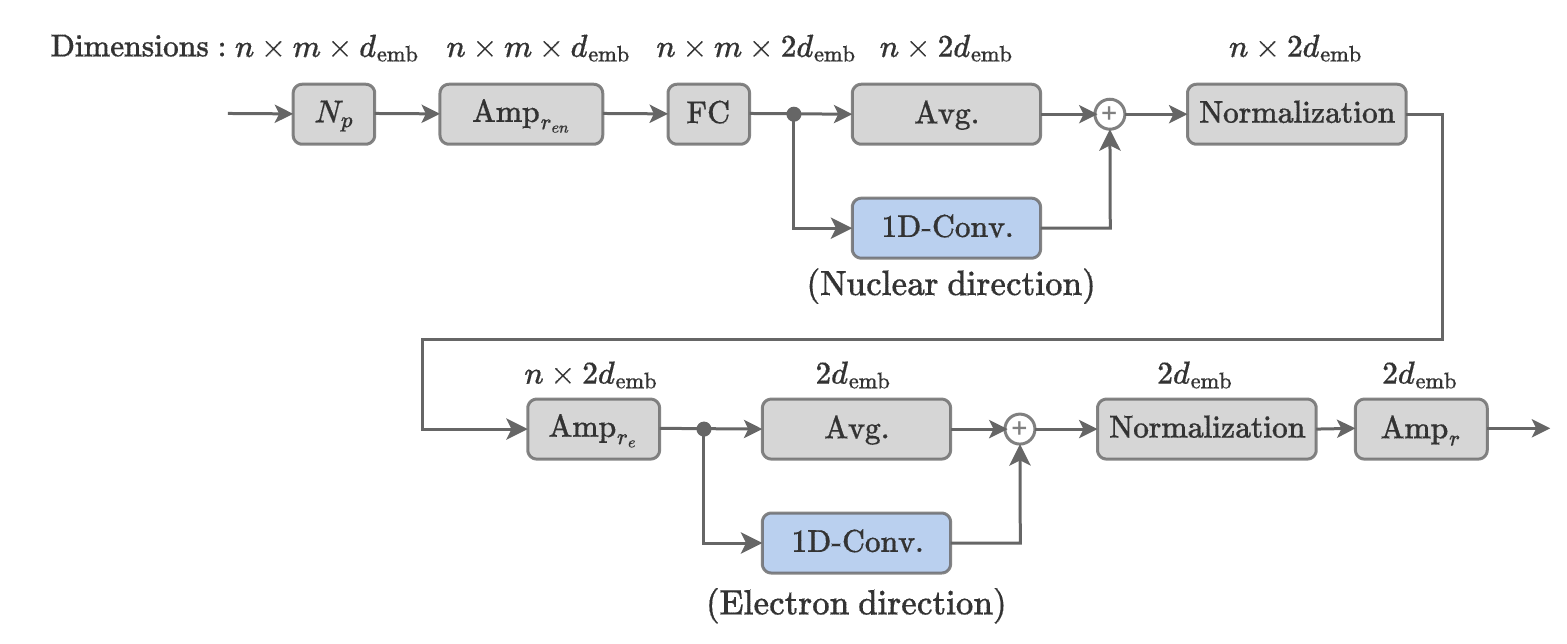}
	\protect{\caption{The aggregation module transforming the Quantum Transformer output matrix  $\by^{(L)}$ ($n\times m \times \demb$) into an $2\demb$-dimensional feature vector. The process applies amplification modules $N_p$ and $\textup{Amp}_{r_\textup{en}}$, followed by an FC module yielding $n\times m \times (2\demb)$ features. These are processed along the nuclear dimension direction by Avg. and 1D-Conv. modules, summed into $n\times (2\demb)$ features, and adjusted by $\textup{Amp}_{r_\textup{e}}$ using $r_{\textup{e}}$ (averaged value of $r_{\textup{en}}$). Further Avg. and 1D-Conv. modules along the electron dimension direction, followed by a normalization and  $\textup{Amp}_{r}$, produce the final vector.\label{Figure:aggregation}}}
\end{figure*}

Figure~\ref{Figure:attention}(a) provides a detailed description of the tokenization module and Quantum Transformer module. Following a token preparation approach similar to that in Ref.~\cite{glehn:2023a:psiformer}, each $\demb$-dimensional feature input vector $\bx$ in the token matrix is derived from a vector concatenating the relative positions of electrons to each nucleus ($p^{\textup{in}}_{\textup{en}}$) and their initial distances ($r^{\textup{in}}_{\textup{en}}$), processed through a fully-connected (FC) layer with trainable weights $W_{\textup{en}}$. 
The computational steps to obtain $p^{\textup{in}}_{\textup{en}}$ and $r^{\textup{in}}_{\textup{en}}$ are outlined as follows:
\begin{itemize}
    \item \textit{Input}: The input consists of the positions of individual atoms (nuclear coordinates) and electron identifiers (atomic orbital assignments). For example, in $\textup{BeH}_{\textup{2}}$, the 3D coordinates of the Hydrogen (H) atoms are symmetrically positioned along the $z$-axis at $(0,0,-2.5)$ and $(0,0,2.5)$, with the Beryllium (Be) atom at the origin $(0, 0, 0)$.
    The nucleus position is an $m\times 3$ matrix with $m=3$, where the first and last rows correspond to the two H atoms.
    Electron identifiers are assigned as follows: [1] (1s) for the first H, [1] (1s), [2] (1s), [3] (2s), and [4] (2s) for Be, and [1] (1s) for the second H.
    
    \item \textit{Embedding (1)}: Each electron identifier is converted into a one-hot vector of dimension equal to the maximum identifier (4 in the $\textup{BeH}_{\textup{2}}$ example). Thus, all $n$ electron identifiers are represented as an $n\times 4$ binary matrix. This matrix is then multiplied by a $4\times 3$ trainable weight matrix to produce an $n \times 3$ matrix of electron relative positions. These positions are relative to their respective atoms, and the trainable weights ensure that electron positions adapt to changes in the molecular structure.
    
    \item \textit{Adder (2)}: The absolute position of each electron is computed by adding its relative position to the position of its associated atom, resulting in an $n\times 3$ matrix.
    
    \item \textit{Subtraction (3)}: The relative positions between all electrons and all nuclei are calculated. For each nucleus, an $n\times 3$ matrix is obtained by subtracting the absolute positions of all electrons from that nucleus's position, yielding an output matrix $p^{\textup{in}}_{\textup{en}}$ of shape $n\times m \times 3$.
    
    \item \textit{Concatenation (4)}: The relative distance matrix $r^{\textup{in}}_{\textup{en}}$, with shape $n\times m \times 1$, is derived from $p^{\textup{in}}_{\textup{en}}$ where each element in $r^{\textup{in}}_{\textup{en}}$ is the Euclidean norm of each position vector. Concatenating $p^{\textup{in}}_{\textup{en}}$ with $r^{\textup{in}}_{\textup{en}}$ produces an $n\times m \times 4$ matrix. 
\end{itemize}

\subsubsection*{Quantum Transformer}
The architecture of the Quantum Transformer component in MQT is depicted in the lower panel of Fig.~\ref{Figure:attention}(a), consisting of $L$ repeated layers. In each layer, the input features derived from the classical representation of the molecule are processed through the $\textup{arccos}$, $\textup{tanh}$, and amplification $N_p$ modules. For each electron, this generates an $m\times \demb$ features matrix corresponding to $m$ tokens $\bx_j$. Each $\demb$-dimensional feature vector of token $\bx_j$ is embedded into a quantum state $\ket{\psi_{j,0}}$ via an angle embedding layer using $\demb$ qubits. Query, Key, and Value transformations, implemented with quantum ansatzes, are then applied to update the states $\ket{\psi_{j,k-1}}\to\ket{\psi_{j,k}}$ $(k=1,\ldots,m)$ through a quantum self-attention mechanism. The final quantum state $\ket{\psi_{j,m}}$ is measured in the computational basis and converted back into classical features of dimension $m \times \demb$, yielding an $n\times m \times \demb$ feature matrix for $n$ blocks of $n$ electrons. These features are added to the input features via a residual connection and normalization to produce the input for the next layer.

The detailed of the update $\ket{\psi_{j,k-1}}\to\ket{\psi_{j,k}}$ is shown in Fig.~\ref{Figure:attention}(b). 
For an input token $\bx_j$, angle embedding is performed using a $\textup{R}_Y$ rotation gate ($\textup{Embed.}$) on a second auxiliary token register. This is followed by a Query transformation and the adjoint of Key transformation, both constructed with a single layer of strongly entangling  (StrEnt) ansatz. Then the adjoint matrix of the angle embedding ($\textup{Embed.}^{\dagger}$) for token $\bx_k$ is applied to compute the Hadamard product of the Query and Key, which resembles the single-head attention. This product is then transformed into a 1-qubit attention representation using a 2-bond matrix product state (MPS) ansatz implemented with trainable $\textup{R}_Y$ and an additional ancilla qubit.
Next, a Value transformation for $\bx_k$ is applied using the first auxiliary token register, involving the angle embedding (Embed.) and a six-layer StrEnt ansatz.
Finally, a controlled SWAP gate updates the features based on the attention representation.

\subsubsection*{Aggregation}
The outputs of all Quantum Transformer modules are summarized into a feature matrix $\by^{(L)}$ of dimension $n\times m \times \demb$, 
which is then transformed into a $2\demb$-dimensional feature vector via an aggregation module.
We illustrate the aggregation module in Fig.~\ref{Figure:aggregation}. Initially, the amplification modules $N_p$ and $\textup{Amp}_{r_\textup{en}}$ are applied to $\by^{(L)}$ without altering its dimension.
Each element $y_{ijk}$  ($1\leq i \leq n$, $1\leq j \leq m$, $1\leq k \leq \demb$) in $\by^{(L)}$ is transformed as:
\begin{align}
    y_{ijk} \to e^{-r_{\textup{en}}(i,j,0)} N_p  y_{ijk},
\end{align}
where $N_p$ is the proton number, $r_{\textup{en}}$ is the $n\times m \times 1$ electron-nucleus relative distance matrix with elements $r_{\textup{en}}(i,j,0)$ ($1\leq i \leq n$, $1\leq j \leq m$).
The relative distance $r_{\textup{en}}$ is updated from the initial distance $r^{\textup{in}}_{\textup{en}}$ using the attention scheme through the following inversion:
\begin{align}
    [p^{\textup{out}}_{\textup{en}}, r^{\textup{out}}_{\textup{en}}]& = \left( W^\top_{\textup{en}} \times W_{\textup{en}}\right)^{-1} \times W^\top_{\textup{en}} \times \by^{(L)},\\
    r_{\textup{en}}&=\textup{std}(r^{\textup{in}}_{\textup{en}})\times \dfrac{r^{\textup{out}}_{\textup{en}}-\textup{mean}(r^{\textup{out}}_{\textup{en}})}{\textup{std}(r^{\textup{out}}_{\textup{en}})} + \textup{mean}(r^{\textup{in}}_{\textup{en}})\label{eqn:renc:update},
\end{align}
where $\textup{mean}(A)$ and $\textup{std}(A)$ denote for the scalar mean and standard deviation of all elements in matrix $A$, and $r^{\textup{in}}_{\textup{en}}$ and $r^{\textup{out}}_{\textup{en}}$ share the $n\times m \times 1$ dimension. Here, $\times, +, -$ are  broadcast operators applied element-wise to $r^{\textup{out}}_{\textup{en}}$.

The output of $\textup{Amp}_{r_\textup{en}}$ with dimension $n\times m \times \demb$ is processed by an FC module with trainable weights to obtain features of dimension $n\times m \times (2\demb)$. 
These features are then separately processed along the nuclear dimension direction (second axis) by an averaging (Avg.) module and a one-dimensional convolutional (1D-Conv.) module (kernel size 2, stride 2), and the results are summed to produce an $n\times (2\demb)$ feature matrix.
This matrix is normalized and adjusted by the amplification module $\textup{Amp}_{r_{\textup{e}}}$, where $r_{\textup{e}}$ (dimension $n$) is the average $r_{\textup{en}}$ across its second dimension. The amplification module $\textup{Amp}_{r_{\textup{e}}}$ multiplies each element at the index $(i,k)$ ($1\leq i \leq n$, $1\leq k \leq 2\demb$) of the feature matrix to $e^{-r_{\textup{e}}(i)}$ where $r_e(i)$ is the $i$-element in $r_{\textup{e}}$.
The output is then separately processed by Avg. and 1D-Conv. modules along the electron dimension direction (the first axis), summed to form a $2\demb$-dimensional feature vector, and further normalized and amplified by $\textup{Amp}_{r}$. Here $r$ is the scalar average of $r_{\textup{e}}$, and each element of the vector is multiplied by $e^{-r}$.

\bibliography{main.bib}

\providecommand{\noopsort}[1]{}\providecommand{\singleletter}[1]{#1}%
\begin{thebibliography}{50}%
\makeatletter
\providecommand \@ifxundefined [1]{%
 \@ifx{#1\undefined}
}%
\providecommand \@ifnum [1]{%
 \ifnum #1\expandafter \@firstoftwo
 \else \expandafter \@secondoftwo
 \fi
}%
\providecommand \@ifx [1]{%
 \ifx #1\expandafter \@firstoftwo
 \else \expandafter \@secondoftwo
 \fi
}%
\providecommand \natexlab [1]{#1}%
\providecommand \enquote  [1]{``#1''}%
\providecommand \bibnamefont  [1]{#1}%
\providecommand \bibfnamefont [1]{#1}%
\providecommand \citenamefont [1]{#1}%
\providecommand \href@noop [0]{\@secondoftwo}%
\providecommand \href [0]{\begingroup \@sanitize@url \@href}%
\providecommand \@href[1]{\@@startlink{#1}\@@href}%
\providecommand \@@href[1]{\endgroup#1\@@endlink}%
\providecommand \@sanitize@url [0]{\catcode `\\12\catcode `\$12\catcode
  `\&12\catcode `\#12\catcode `\^12\catcode `\_12\catcode `\%12\relax}%
\providecommand \@@startlink[1]{}%
\providecommand \@@endlink[0]{}%
\providecommand \url  [0]{\begingroup\@sanitize@url \@url }%
\providecommand \@url [1]{\endgroup\@href {#1}{\urlprefix }}%
\providecommand \urlprefix  [0]{URL }%
\providecommand \Eprint [0]{\href }%
\providecommand \doibase [0]{https://doi.org/}%
\providecommand \selectlanguage [0]{\@gobble}%
\providecommand \bibinfo  [0]{\@secondoftwo}%
\providecommand \bibfield  [0]{\@secondoftwo}%
\providecommand \translation [1]{[#1]}%
\providecommand \BibitemOpen [0]{}%
\providecommand \bibitemStop [0]{}%
\providecommand \bibitemNoStop [0]{.\EOS\space}%
\providecommand \EOS [0]{\spacefactor3000\relax}%
\providecommand \BibitemShut  [1]{\csname bibitem#1\endcsname}%
\let\auto@bib@innerbib\@empty
\bibitem [{\citenamefont {Vaswani}\ \emph {et~al.}(2017)\citenamefont
  {Vaswani}, \citenamefont {Shazeer}, \citenamefont {Parmar}, \citenamefont
  {Uszkoreit}, \citenamefont {Jones}, \citenamefont {Gomez}, \citenamefont
  {Kaiser},\ and\ \citenamefont {Polosukhin}}]{attention:2017:NIPS}%
  \BibitemOpen
  \bibfield  {author} {\bibinfo {author} {\bibfnamefont {A.}~\bibnamefont
  {Vaswani}}, \bibinfo {author} {\bibfnamefont {N.}~\bibnamefont {Shazeer}},
  \bibinfo {author} {\bibfnamefont {N.}~\bibnamefont {Parmar}}, \bibinfo
  {author} {\bibfnamefont {J.}~\bibnamefont {Uszkoreit}}, \bibinfo {author}
  {\bibfnamefont {L.}~\bibnamefont {Jones}}, \bibinfo {author} {\bibfnamefont
  {A.~N.}\ \bibnamefont {Gomez}}, \bibinfo {author} {\bibfnamefont {L.~u.}\
  \bibnamefont {Kaiser}},\ and\ \bibinfo {author} {\bibfnamefont
  {I.}~\bibnamefont {Polosukhin}},\ }in\ \href
  {https://proceedings.neurips.cc/paper_files/paper/2017/file/3f5ee243547dee91fbd053c1c4a845aa-Paper.pdf}
  {\emph {\bibinfo {booktitle} {Advances in Neural Information Processing
  Systems}}},\ Vol.~\bibinfo {volume} {30}\ (\bibinfo  {publisher} {Curran
  Associates, Inc.},\ \bibinfo {year} {2017})\BibitemShut {NoStop}%
\bibitem [{\citenamefont {Dunjko}\ \emph {et~al.}(2016)\citenamefont {Dunjko},
  \citenamefont {Taylor},\ and\ \citenamefont
  {Briegel}}]{dunjko:2016:prl:qmlenhanced}%
  \BibitemOpen
  \bibfield  {author} {\bibinfo {author} {\bibfnamefont {V.}~\bibnamefont
  {Dunjko}}, \bibinfo {author} {\bibfnamefont {J.~M.}\ \bibnamefont {Taylor}},\
  and\ \bibinfo {author} {\bibfnamefont {H.~J.}\ \bibnamefont {Briegel}},\
  }\href {https://doi.org/10.1103/PhysRevLett.117.130501} {\bibfield  {journal}
  {\bibinfo  {journal} {Phys. Rev. Lett.}\ }\textbf {\bibinfo {volume} {117}},\
  \bibinfo {pages} {130501} (\bibinfo {year} {2016})}\BibitemShut {NoStop}%
\bibitem [{\citenamefont {Biamonte}\ \emph {et~al.}(2017)\citenamefont
  {Biamonte}, \citenamefont {Wittek}, \citenamefont {Pancotti}, \citenamefont
  {Rebentrost}, \citenamefont {Wiebe},\ and\ \citenamefont
  {Lloyd}}]{biamonte:2017:QML}%
  \BibitemOpen
  \bibfield  {author} {\bibinfo {author} {\bibfnamefont {J.}~\bibnamefont
  {Biamonte}}, \bibinfo {author} {\bibfnamefont {P.}~\bibnamefont {Wittek}},
  \bibinfo {author} {\bibfnamefont {N.}~\bibnamefont {Pancotti}}, \bibinfo
  {author} {\bibfnamefont {P.}~\bibnamefont {Rebentrost}}, \bibinfo {author}
  {\bibfnamefont {N.}~\bibnamefont {Wiebe}},\ and\ \bibinfo {author}
  {\bibfnamefont {S.}~\bibnamefont {Lloyd}},\ }\href
  {https://doi.org/10.1038/nature23474} {\bibfield  {journal} {\bibinfo
  {journal} {Nature}\ }\textbf {\bibinfo {volume} {549}},\ \bibinfo {pages}
  {195} (\bibinfo {year} {2017})}\BibitemShut {NoStop}%
\bibitem [{\citenamefont {Dunjko}\ and\ \citenamefont
  {Wittek}(2020)}]{dunjko:2020:nonreview}%
  \BibitemOpen
  \bibfield  {author} {\bibinfo {author} {\bibfnamefont {V.}~\bibnamefont
  {Dunjko}}\ and\ \bibinfo {author} {\bibfnamefont {P.}~\bibnamefont
  {Wittek}},\ }\href {https://doi.org/10.22331/qv-2020-03-17-32} {\bibfield
  {journal} {\bibinfo  {journal} {{Quantum Views}}\ }\textbf {\bibinfo {volume}
  {4}},\ \bibinfo {pages} {32} (\bibinfo {year} {2020})}\BibitemShut {NoStop}%
\bibitem [{\citenamefont {Schuld}\ and\ \citenamefont
  {Petruccione}(2021)}]{schuld:2021:qmlbook}%
  \BibitemOpen
  \bibfield  {author} {\bibinfo {author} {\bibfnamefont {M.}~\bibnamefont
  {Schuld}}\ and\ \bibinfo {author} {\bibfnamefont {F.}~\bibnamefont
  {Petruccione}},\ }\href {https://doi.org/10.1007/978-3-030-83098-4} {\emph
  {\bibinfo {title} {Machine Learning with Quantum Computers}}}\ (\bibinfo
  {publisher} {Springer International Publishing},\ \bibinfo {year}
  {2021})\BibitemShut {NoStop}%
\bibitem [{\citenamefont {Havl{\'{\i}}{\v{c}}ek}\ \emph
  {et~al.}(2019)\citenamefont {Havl{\'{\i}}{\v{c}}ek}, \citenamefont
  {C{\'{o}}rcoles}, \citenamefont {Temme}, \citenamefont {Harrow},
  \citenamefont {Kandala}, \citenamefont {Chow},\ and\ \citenamefont
  {Gambetta}}]{halvlicek:2019:supervised}%
  \BibitemOpen
  \bibfield  {author} {\bibinfo {author} {\bibfnamefont {V.}~\bibnamefont
  {Havl{\'{\i}}{\v{c}}ek}}, \bibinfo {author} {\bibfnamefont {A.~D.}\
  \bibnamefont {C{\'{o}}rcoles}}, \bibinfo {author} {\bibfnamefont
  {K.}~\bibnamefont {Temme}}, \bibinfo {author} {\bibfnamefont {A.~W.}\
  \bibnamefont {Harrow}}, \bibinfo {author} {\bibfnamefont {A.}~\bibnamefont
  {Kandala}}, \bibinfo {author} {\bibfnamefont {J.~M.}\ \bibnamefont {Chow}},\
  and\ \bibinfo {author} {\bibfnamefont {J.~M.}\ \bibnamefont {Gambetta}},\
  }\href {https://doi.org/10.1038/s41586-019-0980-2} {\bibfield  {journal}
  {\bibinfo  {journal} {Nature}\ }\textbf {\bibinfo {volume} {567}},\ \bibinfo
  {pages} {209} (\bibinfo {year} {2019})}\BibitemShut {NoStop}%
\bibitem [{\citenamefont {Schuld}\ and\ \citenamefont
  {Killoran}(2019)}]{schuld:2019:feature}%
  \BibitemOpen
  \bibfield  {author} {\bibinfo {author} {\bibfnamefont {M.}~\bibnamefont
  {Schuld}}\ and\ \bibinfo {author} {\bibfnamefont {N.}~\bibnamefont
  {Killoran}},\ }\href {https://doi.org/10.1103/PhysRevLett.122.040504}
  {\bibfield  {journal} {\bibinfo  {journal} {Phys. Rev. Lett.}\ }\textbf
  {\bibinfo {volume} {122}},\ \bibinfo {pages} {040504} (\bibinfo {year}
  {2019})}\BibitemShut {NoStop}%
\bibitem [{\citenamefont {Sipio}\ \emph {et~al.}(2021)\citenamefont {Sipio},
  \citenamefont {Huang}, \citenamefont {Chen}, \citenamefont {Mangini},\ and\
  \citenamefont {Worring}}]{disipio:2021:QNLP}%
  \BibitemOpen
  \bibfield  {author} {\bibinfo {author} {\bibfnamefont {R.~D.}\ \bibnamefont
  {Sipio}}, \bibinfo {author} {\bibfnamefont {J.-H.}\ \bibnamefont {Huang}},
  \bibinfo {author} {\bibfnamefont {S.~Y.-C.}\ \bibnamefont {Chen}}, \bibinfo
  {author} {\bibfnamefont {S.}~\bibnamefont {Mangini}},\ and\ \bibinfo {author}
  {\bibfnamefont {M.}~\bibnamefont {Worring}},\ }\href
  {https://doi.org/10.48550/arxiv.2110.06510} {\bibinfo {title} {The dawn of
  quantum natural language processing}} (\bibinfo {year} {2021})\BibitemShut
  {NoStop}%
\bibitem [{\citenamefont {Li}\ \emph {et~al.}(2024)\citenamefont {Li},
  \citenamefont {Zhao},\ and\ \citenamefont {Wang}}]{li:2024:quasn}%
  \BibitemOpen
  \bibfield  {author} {\bibinfo {author} {\bibfnamefont {G.}~\bibnamefont
  {Li}}, \bibinfo {author} {\bibfnamefont {X.}~\bibnamefont {Zhao}},\ and\
  \bibinfo {author} {\bibfnamefont {X.}~\bibnamefont {Wang}},\ }\href@noop {}
  {\bibfield  {journal} {\bibinfo  {journal} {Sci. China Inf. Sci.}\ }\textbf
  {\bibinfo {volume} {67}},\ \bibinfo {pages} {142501} (\bibinfo {year}
  {2024})}\BibitemShut {NoStop}%
\bibitem [{\citenamefont {Xue}\ \emph {et~al.}(2024)\citenamefont {Xue},
  \citenamefont {Chen}, \citenamefont {Zhuang}, \citenamefont {Wang},
  \citenamefont {Sun}, \citenamefont {Wang}, \citenamefont {Liu}, \citenamefont
  {Wu}, \citenamefont {Wang},\ and\ \citenamefont
  {Guo}}]{xue:2024:quantumvision}%
  \BibitemOpen
  \bibfield  {author} {\bibinfo {author} {\bibfnamefont {C.}~\bibnamefont
  {Xue}}, \bibinfo {author} {\bibfnamefont {Z.-Y.}\ \bibnamefont {Chen}},
  \bibinfo {author} {\bibfnamefont {X.-N.}\ \bibnamefont {Zhuang}}, \bibinfo
  {author} {\bibfnamefont {Y.-J.}\ \bibnamefont {Wang}}, \bibinfo {author}
  {\bibfnamefont {T.-P.}\ \bibnamefont {Sun}}, \bibinfo {author} {\bibfnamefont
  {J.-C.}\ \bibnamefont {Wang}}, \bibinfo {author} {\bibfnamefont {H.-Y.}\
  \bibnamefont {Liu}}, \bibinfo {author} {\bibfnamefont {Y.-C.}\ \bibnamefont
  {Wu}}, \bibinfo {author} {\bibfnamefont {Z.-L.}\ \bibnamefont {Wang}},\ and\
  \bibinfo {author} {\bibfnamefont {G.-P.}\ \bibnamefont {Guo}},\ }\bibfield
  {journal} {\bibinfo  {journal} {arXiv}\ }\href
  {https://doi.org/10.48550/arxiv.2402.18940} {10.48550/arxiv.2402.18940}
  (\bibinfo {year} {2024})\BibitemShut {NoStop}%
\bibitem [{\citenamefont {Giovannetti}\ \emph {et~al.}(2008)\citenamefont
  {Giovannetti}, \citenamefont {Lloyd},\ and\ \citenamefont
  {Maccone}}]{lloyd:2008:QRAM}%
  \BibitemOpen
  \bibfield  {author} {\bibinfo {author} {\bibfnamefont {V.}~\bibnamefont
  {Giovannetti}}, \bibinfo {author} {\bibfnamefont {S.}~\bibnamefont {Lloyd}},\
  and\ \bibinfo {author} {\bibfnamefont {L.}~\bibnamefont {Maccone}},\ }\href
  {https://doi.org/10.1103/PhysRevLett.100.160501} {\bibfield  {journal}
  {\bibinfo  {journal} {Phys. Rev. Lett.}\ }\textbf {\bibinfo {volume} {100}},\
  \bibinfo {pages} {160501} (\bibinfo {year} {2008})}\BibitemShut {NoStop}%
\bibitem [{\citenamefont {Cherrat}\ \emph {et~al.}(2024)\citenamefont
  {Cherrat}, \citenamefont {Kerenidis}, \citenamefont {Mathur}, \citenamefont
  {Landman}, \citenamefont {Strahm},\ and\ \citenamefont
  {Li}}]{cherrat:2024:quantumvision}%
  \BibitemOpen
  \bibfield  {author} {\bibinfo {author} {\bibfnamefont {E.~A.}\ \bibnamefont
  {Cherrat}}, \bibinfo {author} {\bibfnamefont {I.}~\bibnamefont {Kerenidis}},
  \bibinfo {author} {\bibfnamefont {N.}~\bibnamefont {Mathur}}, \bibinfo
  {author} {\bibfnamefont {J.}~\bibnamefont {Landman}}, \bibinfo {author}
  {\bibfnamefont {M.}~\bibnamefont {Strahm}},\ and\ \bibinfo {author}
  {\bibfnamefont {Y.~Y.}\ \bibnamefont {Li}},\ }\href
  {https://doi.org/10.22331/q-2024-02-22-1265} {\bibfield  {journal} {\bibinfo
  {journal} {{Quantum}}\ }\textbf {\bibinfo {volume} {8}},\ \bibinfo {pages}
  {1265} (\bibinfo {year} {2024})}\BibitemShut {NoStop}%
\bibitem [{\citenamefont {Smaldone}\ \emph {et~al.}(2025)\citenamefont
  {Smaldone}, \citenamefont {Shee}, \citenamefont {Kyro}, \citenamefont
  {Farag}, \citenamefont {Chandani}, \citenamefont {Kyoseva},\ and\
  \citenamefont {Batista}}]{smaldone:2025:hybridtrans}%
  \BibitemOpen
  \bibfield  {author} {\bibinfo {author} {\bibfnamefont {A.~M.}\ \bibnamefont
  {Smaldone}}, \bibinfo {author} {\bibfnamefont {Y.}~\bibnamefont {Shee}},
  \bibinfo {author} {\bibfnamefont {G.~W.}\ \bibnamefont {Kyro}}, \bibinfo
  {author} {\bibfnamefont {M.~H.}\ \bibnamefont {Farag}}, \bibinfo {author}
  {\bibfnamefont {Z.}~\bibnamefont {Chandani}}, \bibinfo {author}
  {\bibfnamefont {E.}~\bibnamefont {Kyoseva}},\ and\ \bibinfo {author}
  {\bibfnamefont {V.~S.}\ \bibnamefont {Batista}},\ }\bibfield  {journal}
  {\bibinfo  {journal} {arXiv}\ }\href
  {https://doi.org/10.48550/arxiv.2502.19214} {10.48550/arxiv.2502.19214}
  (\bibinfo {year} {2025})\BibitemShut {NoStop}%
\bibitem [{\citenamefont {Zheng}\ \emph {et~al.}(2023)\citenamefont {Zheng},
  \citenamefont {Gao},\ and\ \citenamefont {Miao}}]{zheng:2023:QSAN}%
  \BibitemOpen
  \bibfield  {author} {\bibinfo {author} {\bibfnamefont {J.}~\bibnamefont
  {Zheng}}, \bibinfo {author} {\bibfnamefont {Q.}~\bibnamefont {Gao}},\ and\
  \bibinfo {author} {\bibfnamefont {Z.}~\bibnamefont {Miao}},\ }in\ \href
  {https://doi.org/10.1109/smc53992.2023.10393989} {\emph {\bibinfo {booktitle}
  {2023 IEEE International Conference on Systems, Man, and Cybernetics
  (SMC)}}}\ (\bibinfo  {publisher} {IEEE},\ \bibinfo {year} {2023})\ p.\
  \bibinfo {pages} {1058–1063}\BibitemShut {NoStop}%
\bibitem [{\citenamefont {Evans}\ \emph {et~al.}(2025)\citenamefont {Evans},
  \citenamefont {Cook}, \citenamefont {Bradshaw},\ and\ \citenamefont
  {LaBorde}}]{evans:2025:qat}%
  \BibitemOpen
  \bibfield  {author} {\bibinfo {author} {\bibfnamefont {E.~N.}\ \bibnamefont
  {Evans}}, \bibinfo {author} {\bibfnamefont {M.}~\bibnamefont {Cook}},
  \bibinfo {author} {\bibfnamefont {Z.~P.}\ \bibnamefont {Bradshaw}},\ and\
  \bibinfo {author} {\bibfnamefont {M.~L.}\ \bibnamefont {LaBorde}},\
  }\bibfield  {journal} {\bibinfo  {journal} {arXiv}\ }\href
  {https://doi.org/10.48550/arxiv.2403.14753} {10.48550/arxiv.2403.14753}
  (\bibinfo {year} {2025})\BibitemShut {NoStop}%
\bibitem [{edi(2023)}]{editorial:2023:qmladv:nature}%
  \BibitemOpen
  \href {https://doi.org/10.1038/s42256-023-00710-9} {\bibfield  {journal}
  {\bibinfo  {journal} {Nat. Mach. Intell.}\ }\textbf {\bibinfo {volume} {5}},\
  \bibinfo {pages} {813–813} (\bibinfo {year} {2023})}\BibitemShut {NoStop}%
\bibitem [{\citenamefont {Peruzzo}\ \emph {et~al.}(2014)\citenamefont
  {Peruzzo}, \citenamefont {McClean}, \citenamefont {Shadbolt}, \citenamefont
  {Yung}, \citenamefont {Zhou}, \citenamefont {Love}, \citenamefont
  {Aspuru-Guzik},\ and\ \citenamefont {O'Brien}}]{peruzzo:2014:VQE}%
  \BibitemOpen
  \bibfield  {author} {\bibinfo {author} {\bibfnamefont {A.}~\bibnamefont
  {Peruzzo}}, \bibinfo {author} {\bibfnamefont {J.}~\bibnamefont {McClean}},
  \bibinfo {author} {\bibfnamefont {P.}~\bibnamefont {Shadbolt}}, \bibinfo
  {author} {\bibfnamefont {M.-H.}\ \bibnamefont {Yung}}, \bibinfo {author}
  {\bibfnamefont {X.-Q.}\ \bibnamefont {Zhou}}, \bibinfo {author}
  {\bibfnamefont {P.~J.}\ \bibnamefont {Love}}, \bibinfo {author}
  {\bibfnamefont {A.}~\bibnamefont {Aspuru-Guzik}},\ and\ \bibinfo {author}
  {\bibfnamefont {J.~L.}\ \bibnamefont {O'Brien}},\ }\href
  {https://doi.org/10.1038/ncomms5213} {\bibfield  {journal} {\bibinfo
  {journal} {Nat. Commun.}\ }\textbf {\bibinfo {volume} {5}},\ \bibinfo {pages}
  {4213} (\bibinfo {year} {2014})}\BibitemShut {NoStop}%
\bibitem [{\citenamefont {Kandala}\ \emph {et~al.}(2017)\citenamefont
  {Kandala}, \citenamefont {Mezzacapo}, \citenamefont {Temme}, \citenamefont
  {Takita}, \citenamefont {Brink}, \citenamefont {Chow},\ and\ \citenamefont
  {Gambetta}}]{kandala:2017:nature:VQE}%
  \BibitemOpen
  \bibfield  {author} {\bibinfo {author} {\bibfnamefont {A.}~\bibnamefont
  {Kandala}}, \bibinfo {author} {\bibfnamefont {A.}~\bibnamefont {Mezzacapo}},
  \bibinfo {author} {\bibfnamefont {K.}~\bibnamefont {Temme}}, \bibinfo
  {author} {\bibfnamefont {M.}~\bibnamefont {Takita}}, \bibinfo {author}
  {\bibfnamefont {M.}~\bibnamefont {Brink}}, \bibinfo {author} {\bibfnamefont
  {J.~M.}\ \bibnamefont {Chow}},\ and\ \bibinfo {author} {\bibfnamefont
  {J.~M.}\ \bibnamefont {Gambetta}},\ }\href
  {https://doi.org/10.1038/nature23879} {\bibfield  {journal} {\bibinfo
  {journal} {Nature}\ }\textbf {\bibinfo {volume} {549}},\ \bibinfo {pages}
  {242} (\bibinfo {year} {2017})}\BibitemShut {NoStop}%
\bibitem [{\citenamefont {Kitaev}(1995)}]{Kitaev:1995:QuantumMA}%
  \BibitemOpen
  \bibfield  {author} {\bibinfo {author} {\bibfnamefont {A.~Y.}\ \bibnamefont
  {Kitaev}},\ }\href
  {https://eccc.weizmann.ac.il/eccc-reports/1996/TR96-003/index.html}
  {\bibfield  {journal} {\bibinfo  {journal} {Electron. Colloquium Comput.
  Complex.}\ }\textbf {\bibinfo {volume} {TR96}} (\bibinfo {year}
  {1995})}\BibitemShut {NoStop}%
\bibitem [{\citenamefont {Nielsen}\ and\ \citenamefont
  {Chuang}(2010)}]{nielsen:2010:quantum}%
  \BibitemOpen
  \bibfield  {author} {\bibinfo {author} {\bibfnamefont {M.~A.}\ \bibnamefont
  {Nielsen}}\ and\ \bibinfo {author} {\bibfnamefont {I.~L.}\ \bibnamefont
  {Chuang}},\ }\href {https://doi.org/10.1017/CBO9780511976667} {\emph
  {\bibinfo {title} {Quantum Computation and Quantum Information}}},\ \bibinfo
  {edition} {10th}\ ed.\ (\bibinfo  {publisher} {Cambridge University Press},\
  \bibinfo {year} {2010})\BibitemShut {NoStop}%
\bibitem [{\citenamefont {Lee}\ \emph {et~al.}(2021)\citenamefont {Lee},
  \citenamefont {Berry}, \citenamefont {Gidney}, \citenamefont {Huggins},
  \citenamefont {McClean}, \citenamefont {Wiebe},\ and\ \citenamefont
  {Babbush}}]{femoco:2021:prxquantum}%
  \BibitemOpen
  \bibfield  {author} {\bibinfo {author} {\bibfnamefont {J.}~\bibnamefont
  {Lee}}, \bibinfo {author} {\bibfnamefont {D.~W.}\ \bibnamefont {Berry}},
  \bibinfo {author} {\bibfnamefont {C.}~\bibnamefont {Gidney}}, \bibinfo
  {author} {\bibfnamefont {W.~J.}\ \bibnamefont {Huggins}}, \bibinfo {author}
  {\bibfnamefont {J.~R.}\ \bibnamefont {McClean}}, \bibinfo {author}
  {\bibfnamefont {N.}~\bibnamefont {Wiebe}},\ and\ \bibinfo {author}
  {\bibfnamefont {R.}~\bibnamefont {Babbush}},\ }\href
  {https://doi.org/10.1103/PRXQuantum.2.030305} {\bibfield  {journal} {\bibinfo
   {journal} {PRX Quantum}\ }\textbf {\bibinfo {volume} {2}},\ \bibinfo {pages}
  {030305} (\bibinfo {year} {2021})}\BibitemShut {NoStop}%
\bibitem [{\citenamefont {Tilly}\ \emph {et~al.}(2022)\citenamefont {Tilly},
  \citenamefont {Chen}, \citenamefont {Cao}, \citenamefont {Picozzi},
  \citenamefont {Setia}, \citenamefont {Li}, \citenamefont {Grant},
  \citenamefont {Wossnig}, \citenamefont {Rungger}, \citenamefont {Booth},\
  and\ \citenamefont {Tennyson}}]{jules:2022:physrep:vqe:review}%
  \BibitemOpen
  \bibfield  {author} {\bibinfo {author} {\bibfnamefont {J.}~\bibnamefont
  {Tilly}}, \bibinfo {author} {\bibfnamefont {H.}~\bibnamefont {Chen}},
  \bibinfo {author} {\bibfnamefont {S.}~\bibnamefont {Cao}}, \bibinfo {author}
  {\bibfnamefont {D.}~\bibnamefont {Picozzi}}, \bibinfo {author} {\bibfnamefont
  {K.}~\bibnamefont {Setia}}, \bibinfo {author} {\bibfnamefont
  {Y.}~\bibnamefont {Li}}, \bibinfo {author} {\bibfnamefont {E.}~\bibnamefont
  {Grant}}, \bibinfo {author} {\bibfnamefont {L.}~\bibnamefont {Wossnig}},
  \bibinfo {author} {\bibfnamefont {I.}~\bibnamefont {Rungger}}, \bibinfo
  {author} {\bibfnamefont {G.~H.}\ \bibnamefont {Booth}},\ and\ \bibinfo
  {author} {\bibfnamefont {J.}~\bibnamefont {Tennyson}},\ }\href
  {https://doi.org/https://doi.org/10.1016/j.physrep.2022.08.003} {\bibfield
  {journal} {\bibinfo  {journal} {Physics Reports}\ }\textbf {\bibinfo {volume}
  {986}},\ \bibinfo {pages} {1} (\bibinfo {year} {2022})},\ \bibinfo {note}
  {the Variational Quantum Eigensolver: a review of methods and best
  practices}\BibitemShut {NoStop}%
\bibitem [{\citenamefont {Gonthier}\ \emph {et~al.}(2022)\citenamefont
  {Gonthier}, \citenamefont {Radin}, \citenamefont {Buda}, \citenamefont
  {Doskocil}, \citenamefont {Abuan},\ and\ \citenamefont
  {Romero}}]{gonthier:2022:VQEblock}%
  \BibitemOpen
  \bibfield  {author} {\bibinfo {author} {\bibfnamefont {J.~F.}\ \bibnamefont
  {Gonthier}}, \bibinfo {author} {\bibfnamefont {M.~D.}\ \bibnamefont {Radin}},
  \bibinfo {author} {\bibfnamefont {C.}~\bibnamefont {Buda}}, \bibinfo {author}
  {\bibfnamefont {E.~J.}\ \bibnamefont {Doskocil}}, \bibinfo {author}
  {\bibfnamefont {C.~M.}\ \bibnamefont {Abuan}},\ and\ \bibinfo {author}
  {\bibfnamefont {J.}~\bibnamefont {Romero}},\ }\href
  {https://doi.org/10.1103/PhysRevResearch.4.033154} {\bibfield  {journal}
  {\bibinfo  {journal} {Phys. Rev. Res.}\ }\textbf {\bibinfo {volume} {4}},\
  \bibinfo {pages} {033154} (\bibinfo {year} {2022})}\BibitemShut {NoStop}%
\bibitem [{\citenamefont {Cervera-Lierta}\ \emph {et~al.}(2021)\citenamefont
  {Cervera-Lierta}, \citenamefont {Kottmann},\ and\ \citenamefont
  {Aspuru-Guzik}}]{cervera:2021:metaVQE}%
  \BibitemOpen
  \bibfield  {author} {\bibinfo {author} {\bibfnamefont {A.}~\bibnamefont
  {Cervera-Lierta}}, \bibinfo {author} {\bibfnamefont {J.~S.}\ \bibnamefont
  {Kottmann}},\ and\ \bibinfo {author} {\bibfnamefont {A.}~\bibnamefont
  {Aspuru-Guzik}},\ }\href {https://doi.org/10.1103/PRXQuantum.2.020329}
  {\bibfield  {journal} {\bibinfo  {journal} {PRX Quantum}\ }\textbf {\bibinfo
  {volume} {2}},\ \bibinfo {pages} {020329} (\bibinfo {year}
  {2021})}\BibitemShut {NoStop}%
\bibitem [{\citenamefont {Ceroni}\ \emph {et~al.}(2023)\citenamefont {Ceroni},
  \citenamefont {Stetina}, \citenamefont {Kieferova}, \citenamefont {Marrero},
  \citenamefont {Arrazola},\ and\ \citenamefont {Wiebe}}]{ceroni2023:gen-gr}%
  \BibitemOpen
  \bibfield  {author} {\bibinfo {author} {\bibfnamefont {J.}~\bibnamefont
  {Ceroni}}, \bibinfo {author} {\bibfnamefont {T.~F.}\ \bibnamefont {Stetina}},
  \bibinfo {author} {\bibfnamefont {M.}~\bibnamefont {Kieferova}}, \bibinfo
  {author} {\bibfnamefont {C.~O.}\ \bibnamefont {Marrero}}, \bibinfo {author}
  {\bibfnamefont {J.~M.}\ \bibnamefont {Arrazola}},\ and\ \bibinfo {author}
  {\bibfnamefont {N.}~\bibnamefont {Wiebe}},\ }\bibfield  {journal} {\bibinfo
  {journal} {arXiv}\ }\href {https://doi.org/10.48550/arXiv.2210.05489}
  {10.48550/arXiv.2210.05489} (\bibinfo {year} {2023})\BibitemShut {NoStop}%
\bibitem [{\citenamefont {Szabo}\ and\ \citenamefont
  {Ostlund}(1996)}]{Szabo1996}%
  \BibitemOpen
  \bibfield  {author} {\bibinfo {author} {\bibfnamefont {A.}~\bibnamefont
  {Szabo}}\ and\ \bibinfo {author} {\bibfnamefont {N.~S.}\ \bibnamefont
  {Ostlund}},\ }\href@noop {} {\emph {\bibinfo {title} {Modern Quantum
  Chemistry: Introduction to Advanced Electronic Structure Theory}}}\ (\bibinfo
   {publisher} {Dover Publications},\ \bibinfo {year} {1996})\BibitemShut
  {NoStop}%
\bibitem [{\citenamefont {Bartlett}\ and\ \citenamefont
  {Musia\l{}}(2007)}]{Bartlett2007}%
  \BibitemOpen
  \bibfield  {author} {\bibinfo {author} {\bibfnamefont {R.~J.}\ \bibnamefont
  {Bartlett}}\ and\ \bibinfo {author} {\bibfnamefont {M.}~\bibnamefont
  {Musia\l{}}},\ }\href {https://doi.org/10.1103/RevModPhys.79.291} {\bibfield
  {journal} {\bibinfo  {journal} {Rev. Mod. Phys.}\ }\textbf {\bibinfo {volume}
  {79}},\ \bibinfo {pages} {291} (\bibinfo {year} {2007})}\BibitemShut
  {NoStop}%
\bibitem [{\citenamefont {M\o{}ller}\ and\ \citenamefont
  {Plesset}(1934)}]{moller:1934}%
  \BibitemOpen
  \bibfield  {author} {\bibinfo {author} {\bibfnamefont {C.}~\bibnamefont
  {M\o{}ller}}\ and\ \bibinfo {author} {\bibfnamefont {M.~S.}\ \bibnamefont
  {Plesset}},\ }\href {https://doi.org/10.1103/PhysRev.46.618} {\bibfield
  {journal} {\bibinfo  {journal} {Phys. Rev.}\ }\textbf {\bibinfo {volume}
  {46}},\ \bibinfo {pages} {618} (\bibinfo {year} {1934})}\BibitemShut
  {NoStop}%
\bibitem [{\citenamefont {Kohn}\ \emph {et~al.}(1996)\citenamefont {Kohn},
  \citenamefont {Becke},\ and\ \citenamefont {Parr}}]{Kohn1996}%
  \BibitemOpen
  \bibfield  {author} {\bibinfo {author} {\bibfnamefont {W.}~\bibnamefont
  {Kohn}}, \bibinfo {author} {\bibfnamefont {A.~D.}\ \bibnamefont {Becke}},\
  and\ \bibinfo {author} {\bibfnamefont {R.~G.}\ \bibnamefont {Parr}},\ }\href
  {https://doi.org/10.1021/jp960669l} {\bibfield  {journal} {\bibinfo
  {journal} {J. Phys. Chem.}\ }\textbf {\bibinfo {volume} {100}},\ \bibinfo
  {pages} {12974–12980} (\bibinfo {year} {1996})}\BibitemShut {NoStop}%
\bibitem [{\citenamefont {Hammond}\ \emph {et~al.}(1994)\citenamefont
  {Hammond}, \citenamefont {Lester},\ and\ \citenamefont
  {Reynolds}}]{Hammond1994}%
  \BibitemOpen
  \bibfield  {author} {\bibinfo {author} {\bibfnamefont {B.~L.}\ \bibnamefont
  {Hammond}}, \bibinfo {author} {\bibfnamefont {W.~A.}\ \bibnamefont {Lester},
  \bibfnamefont {Jr.}},\ and\ \bibinfo {author} {\bibfnamefont {P.~J.}\
  \bibnamefont {Reynolds}},\ }\href@noop {} {\emph {\bibinfo {title} {Monte
  Carlo Methods in Ab Initio Quantum Chemistry}}}\ (\bibinfo  {publisher}
  {World Scientific},\ \bibinfo {year} {1994})\BibitemShut {NoStop}%
\bibitem [{\citenamefont {White}(1992)}]{white:1992:dmrg}%
  \BibitemOpen
  \bibfield  {author} {\bibinfo {author} {\bibfnamefont {S.~R.}\ \bibnamefont
  {White}},\ }\href {https://doi.org/10.1103/PhysRevLett.69.2863} {\bibfield
  {journal} {\bibinfo  {journal} {Phys. Rev. Lett.}\ }\textbf {\bibinfo
  {volume} {69}},\ \bibinfo {pages} {2863} (\bibinfo {year}
  {1992})}\BibitemShut {NoStop}%
\bibitem [{\citenamefont {Seeley}\ \emph {et~al.}(2012)\citenamefont {Seeley},
  \citenamefont {Richard},\ and\ \citenamefont {Love}}]{seeley:2012:transform}%
  \BibitemOpen
  \bibfield  {author} {\bibinfo {author} {\bibfnamefont {J.~T.}\ \bibnamefont
  {Seeley}}, \bibinfo {author} {\bibfnamefont {M.~J.}\ \bibnamefont
  {Richard}},\ and\ \bibinfo {author} {\bibfnamefont {P.~J.}\ \bibnamefont
  {Love}},\ }\href {https://doi.org/10.1063/1.4768229} {\bibfield  {journal}
  {\bibinfo  {journal} {J. Chem. Phys.}\ }\textbf {\bibinfo {volume} {137}},\
  \bibinfo {pages} {224109} (\bibinfo {year} {2012})}\BibitemShut {NoStop}%
\bibitem [{\citenamefont {Azad}(2023)}]{Utkarsh:2023:Chemistry}%
  \BibitemOpen
  \bibfield  {author} {\bibinfo {author} {\bibfnamefont {U.}~\bibnamefont
  {Azad}},\ }\href@noop {} {\bibinfo {title} {Pennylane quantum chemistry
  datasets}},\ \bibinfo {howpublished}
  {\url{https://pennylane.ai/datasets/collection/qchem}} (\bibinfo {year}
  {2023})\BibitemShut {NoStop}%
\bibitem [{\citenamefont {Bergholm}\ and\ \citenamefont
  {et~al.}(2022)}]{bergholm:2022:pennylane}%
  \BibitemOpen
  \bibfield  {author} {\bibinfo {author} {\bibfnamefont {V.}~\bibnamefont
  {Bergholm}}\ and\ \bibinfo {author} {\bibnamefont {et~al.}},\ }\bibfield
  {journal} {\bibinfo  {journal} {arXiv}\ }\href
  {https://doi.org/10.48550/arXiv.1811.04968} {10.48550/arXiv.1811.04968}
  (\bibinfo {year} {2022})\BibitemShut {NoStop}%
\bibitem [{\citenamefont {Loshchilov}\ and\ \citenamefont
  {Hutter}(2019)}]{loshchilov:2018:decoupled}%
  \BibitemOpen
  \bibfield  {author} {\bibinfo {author} {\bibfnamefont {I.}~\bibnamefont
  {Loshchilov}}\ and\ \bibinfo {author} {\bibfnamefont {F.}~\bibnamefont
  {Hutter}},\ }in\ \href {https://openreview.net/forum?id=Bkg6RiCqY7} {\emph
  {\bibinfo {booktitle} {International Conference on Learning
  Representations}}}\ (\bibinfo {year} {2019})\BibitemShut {NoStop}%
\bibitem [{\citenamefont {Recht}\ \emph {et~al.}(2011)\citenamefont {Recht},
  \citenamefont {Re}, \citenamefont {Wright},\ and\ \citenamefont
  {Niu}}]{hogwild:2011:NIPS}%
  \BibitemOpen
  \bibfield  {author} {\bibinfo {author} {\bibfnamefont {B.}~\bibnamefont
  {Recht}}, \bibinfo {author} {\bibfnamefont {C.}~\bibnamefont {Re}}, \bibinfo
  {author} {\bibfnamefont {S.}~\bibnamefont {Wright}},\ and\ \bibinfo {author}
  {\bibfnamefont {F.}~\bibnamefont {Niu}},\ }in\ \href
  {https://proceedings.neurips.cc/paper_files/paper/2011/file/218a0aefd1d1a4be65601cc6ddc1520e-Paper.pdf}
  {\emph {\bibinfo {booktitle} {Advances in Neural Information Processing
  Systems}}},\ Vol.~\bibinfo {volume} {24}\ (\bibinfo  {publisher} {Curran
  Associates, Inc.},\ \bibinfo {year} {2011})\BibitemShut {NoStop}%
\bibitem [{\citenamefont {Wang}\ \emph {et~al.}(2022)\citenamefont {Wang},
  \citenamefont {Ding}, \citenamefont {Gu}, \citenamefont {Li}, \citenamefont
  {Lin}, \citenamefont {Pan}, \citenamefont {Chong},\ and\ \citenamefont
  {Han}}]{hanruiwang:2022:quantumnas}%
  \BibitemOpen
  \bibfield  {author} {\bibinfo {author} {\bibfnamefont {H.}~\bibnamefont
  {Wang}}, \bibinfo {author} {\bibfnamefont {Y.}~\bibnamefont {Ding}}, \bibinfo
  {author} {\bibfnamefont {J.}~\bibnamefont {Gu}}, \bibinfo {author}
  {\bibfnamefont {Z.}~\bibnamefont {Li}}, \bibinfo {author} {\bibfnamefont
  {Y.}~\bibnamefont {Lin}}, \bibinfo {author} {\bibfnamefont {D.~Z.}\
  \bibnamefont {Pan}}, \bibinfo {author} {\bibfnamefont {F.~T.}\ \bibnamefont
  {Chong}},\ and\ \bibinfo {author} {\bibfnamefont {S.}~\bibnamefont {Han}},\
  }in\ \href {https://doi.org/10.1109/HPCA53966.2022.00057} {\emph {\bibinfo
  {booktitle} {The 28th IEEE International Symposium on High-Performance
  Computer Architecture (HPCA-28)}}}\ (\bibinfo {year} {2022})\BibitemShut
  {NoStop}%
\bibitem [{\citenamefont {Fujii}\ \emph {et~al.}(2022)\citenamefont {Fujii},
  \citenamefont {Mizuta}, \citenamefont {Ueda}, \citenamefont {Mitarai},
  \citenamefont {Mizukami},\ and\ \citenamefont
  {Nakagawa}}]{fujii:2022:VQE:divide}%
  \BibitemOpen
  \bibfield  {author} {\bibinfo {author} {\bibfnamefont {K.}~\bibnamefont
  {Fujii}}, \bibinfo {author} {\bibfnamefont {K.}~\bibnamefont {Mizuta}},
  \bibinfo {author} {\bibfnamefont {H.}~\bibnamefont {Ueda}}, \bibinfo {author}
  {\bibfnamefont {K.}~\bibnamefont {Mitarai}}, \bibinfo {author} {\bibfnamefont
  {W.}~\bibnamefont {Mizukami}},\ and\ \bibinfo {author} {\bibfnamefont
  {Y.~O.}\ \bibnamefont {Nakagawa}},\ }\href
  {https://doi.org/10.1103/PRXQuantum.3.010346} {\bibfield  {journal} {\bibinfo
   {journal} {PRX Quantum}\ }\textbf {\bibinfo {volume} {3}},\ \bibinfo {pages}
  {010346} (\bibinfo {year} {2022})}\BibitemShut {NoStop}%
\bibitem [{\citenamefont {Tran}\ \emph
  {et~al.}(2024{\natexlab{a}})\citenamefont {Tran}, \citenamefont {Kikuchi},\
  and\ \citenamefont {Oshima}}]{tran:2024:varQAE}%
  \BibitemOpen
  \bibfield  {author} {\bibinfo {author} {\bibfnamefont {Q.~H.}\ \bibnamefont
  {Tran}}, \bibinfo {author} {\bibfnamefont {S.}~\bibnamefont {Kikuchi}},\ and\
  \bibinfo {author} {\bibfnamefont {H.}~\bibnamefont {Oshima}},\ }\href
  {https://doi.org/10.1103/PhysRevResearch.6.023181} {\bibfield  {journal}
  {\bibinfo  {journal} {Phys. Rev. Res.}\ }\textbf {\bibinfo {volume} {6}},\
  \bibinfo {pages} {023181} (\bibinfo {year} {2024}{\natexlab{a}})}\BibitemShut
  {NoStop}%
\bibitem [{\citenamefont {Tran}\ \emph
  {et~al.}(2024{\natexlab{b}})\citenamefont {Tran}, \citenamefont {Endo},\ and\
  \citenamefont {Oshima}}]{tran:2024:qcurl}%
  \BibitemOpen
  \bibfield  {author} {\bibinfo {author} {\bibfnamefont {Q.~H.}\ \bibnamefont
  {Tran}}, \bibinfo {author} {\bibfnamefont {Y.}~\bibnamefont {Endo}},\ and\
  \bibinfo {author} {\bibfnamefont {H.}~\bibnamefont {Oshima}},\ }\bibfield
  {journal} {\bibinfo  {journal} {arXiv}\ }\href
  {https://doi.org/10.48550/arXiv.2407.02419} {10.48550/arXiv.2407.02419}
  (\bibinfo {year} {2024}{\natexlab{b}})\BibitemShut {NoStop}%
\bibitem [{\citenamefont {Guo}\ \emph {et~al.}(2024)\citenamefont {Guo},
  \citenamefont {Yu}, \citenamefont {Choi}, \citenamefont {Agrawal},
  \citenamefont {Nakaji}, \citenamefont {Aspuru-Guzik},\ and\ \citenamefont
  {Rebentrost}}]{guo2024:qla:transformer}%
  \BibitemOpen
  \bibfield  {author} {\bibinfo {author} {\bibfnamefont {N.}~\bibnamefont
  {Guo}}, \bibinfo {author} {\bibfnamefont {Z.}~\bibnamefont {Yu}}, \bibinfo
  {author} {\bibfnamefont {M.}~\bibnamefont {Choi}}, \bibinfo {author}
  {\bibfnamefont {A.}~\bibnamefont {Agrawal}}, \bibinfo {author} {\bibfnamefont
  {K.}~\bibnamefont {Nakaji}}, \bibinfo {author} {\bibfnamefont
  {A.}~\bibnamefont {Aspuru-Guzik}},\ and\ \bibinfo {author} {\bibfnamefont
  {P.}~\bibnamefont {Rebentrost}},\ }\bibfield  {journal} {\bibinfo  {journal}
  {arXiv}\ }\href {https://doi.org/10.48550/arxiv.2402.16714}
  {10.48550/arxiv.2402.16714} (\bibinfo {year} {2024})\BibitemShut {NoStop}%
\bibitem [{\citenamefont {Liao}\ and\ \citenamefont
  {Ferrie}(2024)}]{liao:2024:gptquantumcomputer}%
  \BibitemOpen
  \bibfield  {author} {\bibinfo {author} {\bibfnamefont {Y.}~\bibnamefont
  {Liao}}\ and\ \bibinfo {author} {\bibfnamefont {C.}~\bibnamefont {Ferrie}},\
  }\bibfield  {journal} {\bibinfo  {journal} {arXiv}\ }\href
  {https://doi.org/10.48550/arxiv.2403.09418} {10.48550/arxiv.2403.09418}
  (\bibinfo {year} {2024})\BibitemShut {NoStop}%
\bibitem [{\citenamefont {Khatri}\ \emph {et~al.}(2024)\citenamefont {Khatri},
  \citenamefont {Matos}, \citenamefont {Coopmans},\ and\ \citenamefont
  {Clark}}]{khatri:2024:quixer:transformer}%
  \BibitemOpen
  \bibfield  {author} {\bibinfo {author} {\bibfnamefont {N.}~\bibnamefont
  {Khatri}}, \bibinfo {author} {\bibfnamefont {G.}~\bibnamefont {Matos}},
  \bibinfo {author} {\bibfnamefont {L.}~\bibnamefont {Coopmans}},\ and\
  \bibinfo {author} {\bibfnamefont {S.}~\bibnamefont {Clark}},\ }\bibfield
  {journal} {\bibinfo  {journal} {arXiv}\ }\href
  {https://doi.org/10.48550/arxiv.2406.04305} {10.48550/arxiv.2406.04305}
  (\bibinfo {year} {2024})\BibitemShut {NoStop}%
\bibitem [{\citenamefont {Chan}\ and\ \citenamefont {Sharma}(2011)}]{Chan2011}%
  \BibitemOpen
  \bibfield  {author} {\bibinfo {author} {\bibfnamefont {G.~K.-L.}\
  \bibnamefont {Chan}}\ and\ \bibinfo {author} {\bibfnamefont {S.}~\bibnamefont
  {Sharma}},\ }\href {https://doi.org/10.1146/annurev-physchem-032210-103338}
  {\bibfield  {journal} {\bibinfo  {journal} {Annu. Rev. Phys. Chem.}\ }\textbf
  {\bibinfo {volume} {62}},\ \bibinfo {pages} {465} (\bibinfo {year}
  {2011})}\BibitemShut {NoStop}%
\bibitem [{\citenamefont {Farhi}\ \emph {et~al.}(2000)\citenamefont {Farhi},
  \citenamefont {Goldstone}, \citenamefont {Gutmann},\ and\ \citenamefont
  {Sipser}}]{farhi:2000:adiabatic}%
  \BibitemOpen
  \bibfield  {author} {\bibinfo {author} {\bibfnamefont {E.}~\bibnamefont
  {Farhi}}, \bibinfo {author} {\bibfnamefont {J.}~\bibnamefont {Goldstone}},
  \bibinfo {author} {\bibfnamefont {S.}~\bibnamefont {Gutmann}},\ and\ \bibinfo
  {author} {\bibfnamefont {M.}~\bibnamefont {Sipser}},\ }\bibfield  {journal}
  {\bibinfo  {journal} {arXiv}\ }\href
  {https://doi.org/10.48550/arXiv.quant-ph/0001106}
  {10.48550/arXiv.quant-ph/0001106} (\bibinfo {year} {2000})\BibitemShut
  {NoStop}%
\bibitem [{\citenamefont {McArdle}\ \emph {et~al.}(2019)\citenamefont
  {McArdle}, \citenamefont {Jones}, \citenamefont {Endo} \emph
  {et~al.}}]{McArdle:2019:imagine}%
  \BibitemOpen
  \bibfield  {author} {\bibinfo {author} {\bibfnamefont {S.}~\bibnamefont
  {McArdle}}, \bibinfo {author} {\bibfnamefont {T.}~\bibnamefont {Jones}},
  \bibinfo {author} {\bibfnamefont {S.}~\bibnamefont {Endo}}, \emph {et~al.},\
  }\href {https://doi.org/10.1038/s41534-019-0187-2} {\bibfield  {journal}
  {\bibinfo  {journal} {npj Quantum Inf.}\ }\textbf {\bibinfo {volume} {5}},\
  \bibinfo {pages} {75} (\bibinfo {year} {2019})}\BibitemShut {NoStop}%
\bibitem [{\citenamefont {Motta}\ \emph {et~al.}(2023)\citenamefont {Motta},
  \citenamefont {Kirby}, \citenamefont {Liepuoniute}, \citenamefont {Sung},
  \citenamefont {Cohn}, \citenamefont {Mezzacapo}, \citenamefont {Klymko},
  \citenamefont {Nguyen}, \citenamefont {Yoshioka},\ and\ \citenamefont
  {Rice}}]{motta:2023:subspace}%
  \BibitemOpen
  \bibfield  {author} {\bibinfo {author} {\bibfnamefont {M.}~\bibnamefont
  {Motta}}, \bibinfo {author} {\bibfnamefont {W.}~\bibnamefont {Kirby}},
  \bibinfo {author} {\bibfnamefont {I.}~\bibnamefont {Liepuoniute}}, \bibinfo
  {author} {\bibfnamefont {K.~J.}\ \bibnamefont {Sung}}, \bibinfo {author}
  {\bibfnamefont {J.}~\bibnamefont {Cohn}}, \bibinfo {author} {\bibfnamefont
  {A.}~\bibnamefont {Mezzacapo}}, \bibinfo {author} {\bibfnamefont
  {K.}~\bibnamefont {Klymko}}, \bibinfo {author} {\bibfnamefont
  {N.}~\bibnamefont {Nguyen}}, \bibinfo {author} {\bibfnamefont
  {N.}~\bibnamefont {Yoshioka}},\ and\ \bibinfo {author} {\bibfnamefont
  {J.~E.}\ \bibnamefont {Rice}},\ }\bibfield  {journal} {\bibinfo  {journal}
  {arXiv}\ }\href {https://doi.org/10.48550/arXiv.2312.00178}
  {10.48550/arXiv.2312.00178} (\bibinfo {year} {2023})\BibitemShut {NoStop}%
\bibitem [{\citenamefont {Kirby}\ \emph {et~al.}(2023)\citenamefont {Kirby},
  \citenamefont {Motta},\ and\ \citenamefont {Mezzacapo}}]{Kirby:2023:lanczos}%
  \BibitemOpen
  \bibfield  {author} {\bibinfo {author} {\bibfnamefont {W.}~\bibnamefont
  {Kirby}}, \bibinfo {author} {\bibfnamefont {M.}~\bibnamefont {Motta}},\ and\
  \bibinfo {author} {\bibfnamefont {A.}~\bibnamefont {Mezzacapo}},\ }\href
  {https://doi.org/10.22331/q-2023-05-23-1018} {\bibfield  {journal} {\bibinfo
  {journal} {Quantum}\ }\textbf {\bibinfo {volume} {7}},\ \bibinfo {pages}
  {1018} (\bibinfo {year} {2023})}\BibitemShut {NoStop}%
\bibitem [{\citenamefont {dvgodoy}(2020)}]{godoy:2020:dlvisuals}%
  \BibitemOpen
  \bibfield  {author} {\bibinfo {author} {\bibnamefont {dvgodoy}},\ }\href@noop
  {} {\bibinfo {title} {dl-visuals}},\ \bibinfo {howpublished}
  {\url{https://github.com/dvgodoy/dl-visuals}} (\bibinfo {year} {2020}),\
  \bibinfo {note} {accessed February 10, 2025}\BibitemShut {NoStop}%
\bibitem [{\citenamefont {von Glehn}\ \emph {et~al.}(2023)\citenamefont {von
  Glehn}, \citenamefont {Spencer},\ and\ \citenamefont
  {Pfau}}]{glehn:2023a:psiformer}%
  \BibitemOpen
  \bibfield  {author} {\bibinfo {author} {\bibfnamefont {I.}~\bibnamefont {von
  Glehn}}, \bibinfo {author} {\bibfnamefont {J.~S.}\ \bibnamefont {Spencer}},\
  and\ \bibinfo {author} {\bibfnamefont {D.}~\bibnamefont {Pfau}},\ }in\ \href
  {https://openreview.net/forum?id=xveTeHVlF7j} {\emph {\bibinfo {booktitle}
  {The Eleventh International Conference on Learning Representations}}}\
  (\bibinfo {year} {2023})\BibitemShut {NoStop}%
\end{thebibliography}%

\section*{Acknowledgements}
\vspace{-0.5cm}
\noindent
The authors acknowledge Shintaro Sato, Koki Chinzei, and Nasa Matsumoto for their fruitful discussions. 
Special thanks are extended to Koki Chinzei for his valuable comments on the design of the MQT model.

\section*{Author contributions}
\vspace{-0.5cm}
\noindent
Y.K. and Q.T. developed the theoretical aspects and research design of this work. 
Y.K. conducted the numerical experiments.
Y.K., Q.T., Y.E., and H.O. contributed to the technical discussions and manuscript writing.

\section*{Competing interests}
\vspace{-0.5cm}
\noindent
The authors declare no competing interests.

\end{document}